\documentclass[reprint, amsmath, amssymb, aps, showkeys]{revtex4-2}
\usepackage{gensymb}
\usepackage{booktabs}
\usepackage{graphicx}
\usepackage{float}
\graphicspath{{./Figures/}}
\usepackage{multirow}
\usepackage{dcolumn}
\usepackage{bm}
\usepackage{times}
\usepackage{algorithm}
\usepackage{algorithmicx}
\usepackage{algpseudocode}
\usepackage{listings}
\usepackage{natbib}
\usepackage{placeins}
\usepackage{verbatim}
\usepackage{mhchem}
\usepackage{array}
\usepackage[explicit]{titlesec}

\usepackage[colorlinks = true,
linkcolor = blue,
urlcolor  = blue,
citecolor = blue,
anchorcolor = blue]{hyperref}

\newcommand{\beginsupplement}{%
  \setcounter{table}{0}
  \renewcommand{\thetable}{S\arabic{table}}%
  \setcounter{figure}{0}
  \renewcommand{\thefigure}{S\arabic{figure}}%
  \setcounter{section}{0}
  \renewcommand{\thesection}{\arabic{section}}
  \renewcommand{\thesubsection}{\thesection.\arabic{subsection}}
  \renewcommand{\thesubsubsection}{\thesubsection.\arabic{subsubsection}}
  \setcounter{equation}{0}
  \renewcommand{\theequation}{S\arabic{equation}}%
  \setcounter{page}{1}
  \renewcommand{\thefootnote}{\fnsymbol{footnote}}
}
\begin{document}

\preprint{APS/123-QED}

\title{Inverse Materials Design by Large Language Model-Assisted Generative Framework}

\author{Yun Hao$^{1,7,\#}$} 
\author{Che Fan$^{3,\#}$} 

\author{Beilin Ye$^2$}
\author{Wenhao Lu$^5$}
\author{Zhen Lu$^5$}
\author{Peilin Zhao$^6$}
\author{Zhifeng Gao$^4$}\email{gaozf@dp.tech}
\author{Qingyao Wu$^1$}\email{qyw@scut.edu.cn}
\author{Yanhui Liu$^5$}\email{yanhui.liu@iphy.ac.cn}
\author{Tongqi Wen$^{2,7}$}\email{tongqwen@hku.hk}

\affiliation{%
		$^1$School of Software Engineering, South China University of Technology, Guangzhou, China
\\
		$^2$Department of Mechanical Engineering, The University of Hong Kong, Hong Kong, China
\\
		$^3$Department of Materials Science and Engineering, City University of Hong Kong, Hong Kong, China
\\
            $^4$DP Technology, Beijing, China
\\
            $^5$Institute of Physics, Chinese Academy of Sciences, Beijing, China
\\
            $^6$Tencent, AI lab, Shenzhen, China
\\
            $^7$AI for Science Institute, Beijing, China
            \\
        $^\#$These authors contributed equally: Yun Hao, Che Fan.   
            }

\date{\today}

\begin{abstract}

Deep generative models hold great promise for inverse materials design, yet their efficiency and accuracy remain constrained by data scarcity and model architecture. Here, we introduce AlloyGAN, a closed-loop framework that integrates Large Language Model (LLM)-assisted text mining with Conditional Generative Adversarial Networks (CGANs) to enhance data diversity and improve inverse design. Taking alloy discovery as a case study, AlloyGAN systematically refines material candidates through iterative screening and experimental validation. For metallic glasses, the framework predicts thermodynamic properties with discrepancies of less than 8\% from experiments, demonstrating its robustness. By bridging generative AI with domain knowledge and validation workflows, AlloyGAN offers a scalable approach to accelerate the discovery of materials with tailored properties, paving the way for broader applications in materials science.

\end{abstract}
\maketitle

\label{sec:intro}

Materials design typically involves two fundamental problems: forward and inverse problems. The forward problem focuses on understanding the relationship between composition, processing conditions, and material properties. This understanding enables researchers to optimize alloy compositions and processing conditions to achieve enhanced performance. Conversely, the inverse problem is more prevalent in material design and poses the question: ``Given the desired material properties, what composition and processing conditions are required to achieve them?'' The inverse problem is particularly challenging for multi-component materials due to the vast composition space and complex interactions among components. Traditional ``trial-and-error'' experimental approaches are often prohibitively time-consuming and cost-ineffective~\cite{raabe2023accelerating} for such problems. Addressing these challenges thus requires innovative approaches to efficiently navigate the composition space and identify optimal solutions for materials design.

To address the high-dimensional challenges in inverse materials design, machine learning (ML) has gained prominence due to its efficiency and cost-effectiveness. Deep neural networks, for example, excel at capturing complex non-linear relationships from diverse data types, including text, tables, and images. These models can uncover latent features from high-dimensional spaces, enabling accurate predictions and insights for previously unseen data. By leveraging vast amounts of accumulated data, ML approaches can dramatically accelerate inverse design, far outpacing traditional experimental methods~\cite{juan2021accelerating}. 

In inverse materials design, supervised ML approaches often rely on compositions, empirical rules, and physical principles as descriptors for classification or regression tasks~\cite{li2020ai,louie2021discovering,merchant2023scaling,raabe2023accelerating}. Once trained, these models explore new materials compositions with desirable properties, such as excellent glass-forming ability (GFA)~\cite{liu2020machine,ward2016general,zhou2021rational}, superior thermal properties~\cite{ward2018machine}, or targeted elastic properties~\cite{xiong2019machine}. While these methods enable rapid screening of materials candidates, they require extensive labeled data, often unavailable due to the limited size of materials property databases. Moreover, the establishment of unified materials databases, particularly for experimental data, is challenged by the complexity of data structures and the diverse nature of characterization features.

\begin{figure*}
	\includegraphics[width=1\textwidth]{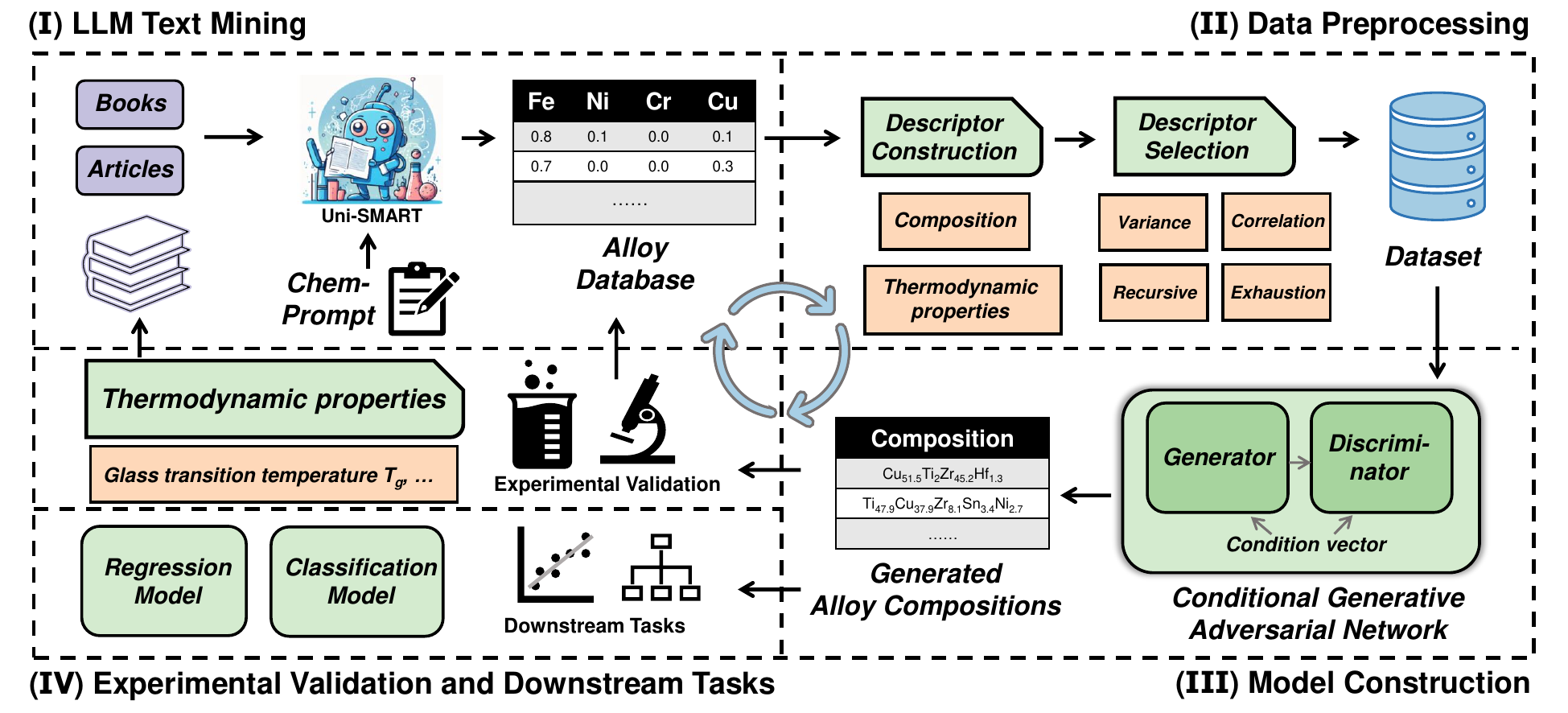}
        \caption{Overview of the AlloyGAN framework. The AlloyGAN framework integrates four key components for targeted composition design and discovering new alloys with specific properties. (I) LLM-Assisted Text Mining: We utilize the Uni-SMART LLM~\textcolor{blue}{[15]} and chemical prompts to extract relevant data from alloy literature. (II) Data Preprocessing: Extracted data is converted into chemical descriptors, with additional descriptor data generated automatically through mathematical formulas to enhance diversity and comprehensiveness. (III) Model Construction: A generative deep learning framework is developed to establish mappings between alloy compositions and properties. The framework employs Conditional Generative Adversarial Networks to generate new alloys that meet specific property requirements. (IV)~Experimental Feedback Loop and Downstream Tasks: Properties of generated alloys are validated through experiments. Verified data is incorporated into the alloy dataset, creating a feedback loop that iteratively enhances model performance and dataset robustness. The framework also supports downstream applications, such as materials classification and property prediction.}
	\label{fig1}
\end{figure*}

Unsupervised ML methods~\cite{xiong2019machine,xiong2020machine} have been applied to materials with a limited number of elements but face significant performance challenges as the number of components exceeds five~\cite{wang2014high,zhang2022recent}. Generative models, including Generative Adversarial Networks (GANs), have also been employed to generate alloy data~\cite{zhou2023generative}. However, many of these methods do not thoroughly analyze the relationships between descriptors, and their generation processes often lack the ability to satisfy specific property requirements. Instead, they tend to generate alloys within predefined categories without tailoring them for target properties. Additionally, the limited accuracy of numerical property predictions in these approaches highlights the need for further verification and methodological improvements. Despite recent progress, these limitations emphasize the necessity of developing more robust frameworks for property-driven alloy design.

To address the challenges outlined above, we propose a conditional generative deep learning AlloyGAN framework for the inverse design of alloys as a case study. 
Our framework begins with leveraging the Large Language Model (LLM) Uni-SMART~\cite{cai2024uni} to streamline database construction for multi-component amorphous alloys. By employing prompt engineering techniques~\cite{zheng2023chatgpt}, we extract relevant data from alloy research papers through automated text mining. This process is further enhanced by generating descriptor data through mathematical formulas, resulting in a more diverse and comprehensive dataset for model training. 
We then develop a generative deep learning model that establishes a direct mapping between alloy composition and material properties, specifically designed for the inverse design of alloys. GANs are utilized to predict alloy compositions that satisfy desired property requirements. To further enhance efficiency and accuracy, we incorporate conditional information into both the generator and discriminator networks to ensure that the generated alloy compositions meet specified property criteria and are validated by the discriminator\cite{mirza2014conditional}. Unlike conventional GANs, which rely on purely random generation, the AlloyGAN framework allows for targeted and property-driven alloy generation. The prediction of AlloyGAN framework is further validated by experiments and the errors on thermophysical properties are within 8\%. Our proposed framework offers significant flexibility and adaptability, making it extendable to other materials applications. By integrating advanced generative techniques and leveraging LLM capabilities, this approach demonstrates its potential as a versatile tool for broader applications in materials science, paving the way for efficient and targeted materials discovery.

\section{Results}

\subsection{Overview of the AlloyGAN Framework}

Fig.~\ref{fig1} shows the AlloyGAN framework developed to enable inverse alloy design for targeted properties. The framework consists of four interconnected components, forming a comprehensive and iterative approach. (I) LLM text mining. The framework begins with text mining assisted by LLMs. Specifically, we utilize carefully designed chemical prompts and Uni-SMART LLM~\cite{cai2024uni} to extract data from books, research papers, and other literature sources. This step ensures the acquisition of relevant information about alloy compositions and properties. (II) Data preprocessing. After text mining, the collected data is processed to create a set of chemical descriptors. These descriptors are further enhanced using mathematical formulas derived from the literature (Supplementary Table~S1), ensuring a diverse and comprehensive dataset that accurately represents the composition-property relationships of alloys. (III) Model construction. The processed dataset is then fed into AlloyGAN. In this step, the generator and discriminator networks are trained adversarially, guided by a condition vector that encodes the desired alloy properties. This process enables the generation of alloy compositions that satisfy the specified property requirements. (IV) Downstream tasks and experimental validation. The generated alloy compositions are validated through experimental synthesis and measurements. The experimentally verified data is then incorporated into the existing database, creating an active learning loop. This feedback mechanism continuously improves the model’s predictive accuracy and expands the alloy dataset. The model constructed in (III) also applies to downstream tasks such as materials classification and property prediction. By integrating these components into a cohesive workflow, the AlloyGAN framework provides an efficient and scalable solution for inverse alloy design, facilitating the discovery of novel materials with tailored properties. In the following, we use multi-component metallic glasses to illustrate the details of each step in the AlloyGAN framework.  

\subsection{Collection and Visualization of Alloy Data}

To construct a comprehensive dataset for inverse alloy design, we leveraged the capabilities of LLMs for data mining and processing, resulting in a dataset of 1,338
entries. This dataset encompasses a diverse range of compositions and thermodynamic properties, including glass-forming temperature~($T_g$), onset crystallization temperature~($T_x$), and liquidus temperature~($T_l$). Using carefully designed prompts, LLMs extracted and synthesized alloy data from diverse sources, focusing on critical descriptors, thermodynamic properties, and compositional details. This automated approach significantly reduced manual effort while improving both the accuracy and breadth of the collected data.

\begin{figure}[h]
	\begin{center}
		\centering
        \includegraphics[width=0.52\textwidth]{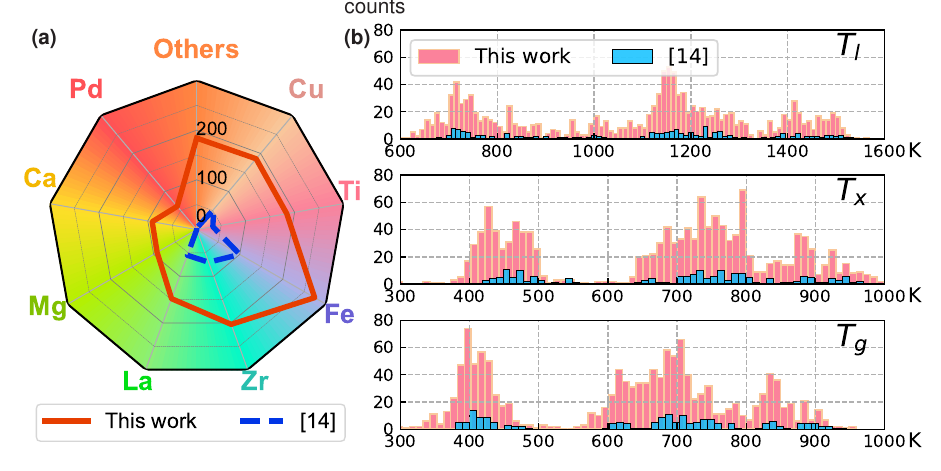}
	\end{center}
	\caption{Expanded dataset comparison and distribution. (a) The dataset collected through LLM-based data mining exhibits broader coverage and enhanced diversity compared to the previous dataset~\cite{zhou2023generative}, highlighting the effectiveness of our approach in capturing a wider range of alloy compositions and properties. (b) Histogram of thermodynamic parameters ($T_l$, $T_x$, $T_g$) for alloys from~\cite{zhou2023generative}, along with the alloy data collected by LLM in this work.}
	\label{element}
\end{figure}

The integration of LLM-driven data mining and descriptor generation streamlined the data acquisition process, resulting in a diverse and comprehensive dataset. As shown in Fig.~\ref{element}(a), our approach surpasses traditional methods that rely on static and limited datasets by enabling the inclusion of alloys with unique or underrepresented compositions and properties.
Fig.~\ref{element}(b) presents the histogram of thermodynamic parameters ($T_l$, $T_x$, and $T_g$) for alloys from \cite{zhou2023generative} (blue bars) and all the alloy data in this work collected by LLM (red bars). This expanded dataset significantly enhances the diversity of thermodynamic properties.

We selected four categories: Cu-, Fe-, Ti-, and Zr-based alloys as our training dataset, referred to as the Cu, Fe, Ti, and Zr datasets. The final training dataset comprised 156 Cu, 257 Fe, 167 Ti, and 213 Zr entries. Compared to datasets used in previous studies~\cite{zhou2023generative}, our dataset significantly improves model stability during training and increases the quality of the generated alloy samples. These advancements enable the AlloyGAN framework to explore a broader distribution of alloy compositions and properties, underscoring the effectiveness of LLMs in creating a larger, more diverse, and reliable dataset for alloy design.

\subsection{Generation of New Alloy Compositions}

\begin{figure*}[ht]
	\begin{center}
		\centering
    \vspace{-25pt}
		\includegraphics[width=1\textwidth]{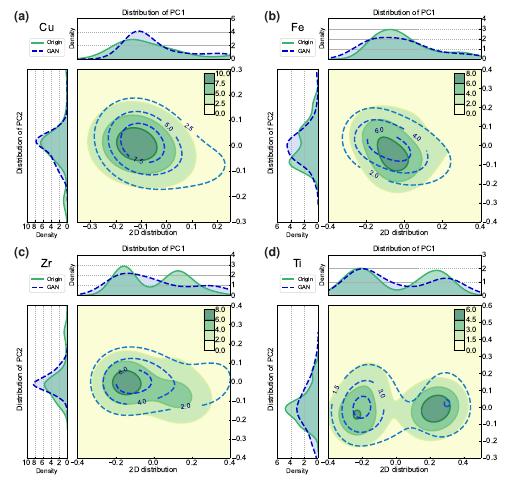}
	\end{center}
    \vspace{-25pt}
	\caption{Comparison of GAN-generated and original alloy compositions across different alloy categories. (a)-(d): Two-dimensional PCA plots illustrating the distributions of GAN-generated and original samples for Cu-, Fe-, Zr-, and Ti-based alloys, respectively. The green solid lines represent the distribution of the original data collected through LLM-assisted data mining, while the blue dashed lines show the distribution of the data generated by the GAN model. The close alignment of the distributions highlights the ability of the GAN model to effectively capture the characteristic features of the original dataset.}
	\label{pca_plots}
\end{figure*}

Using the collected dataset, we trained a GAN model by combining thermodynamic descriptors with alloy compositions. GAN generated 100 new alloy compositions for each alloy category (Cu-, Fe-, Ti-, and Zr-based). To evaluate the similarities and differences between the original and generated datasets, we conducted principal component analysis (PCA)~\cite{yang2004two}. The two-dimensional probability density distributions for both datasets are presented in Fig.~\ref{pca_plots}. The green solid lines represent the distribution of the original data obtained through LLM-assisted data mining, while the blue dashed lines depict the distribution of the GAN-generated data. The results indicate that the GAN successfully replicates the feature-space distribution of the original dataset, reflecting high fidelity. Notably, the PCA analysis reveals distinct patterns in the original dataset: Cu- and Fe-based alloys exhibit single-peak distributions, Ti-based alloys display double-peak distributions, and Zr-based alloys fall between single- and double-peak distributions. Impressively, the GAN-generated dataset captures these unique characteristics, accurately reflecting the diversity and complexity inherent in the original data.

The close alignment between the PCA distributions of the original and GAN-generated datasets underscores the reliability and representativeness of the generated alloy compositions. This validation demonstrates the ability of GAN to effectively capture and reproduce the essential features of various alloy systems. By reproducing the critical characteristics of the original data, these results establish a foundation for accurate inverse alloy design. 

\subsection{Inverse Alloy Design}

\begin{figure*}[ht]
	\begin{center}
		\centering
          \vspace{-30pt}\hspace{1.2cm}\includegraphics[width=1.05\textwidth]{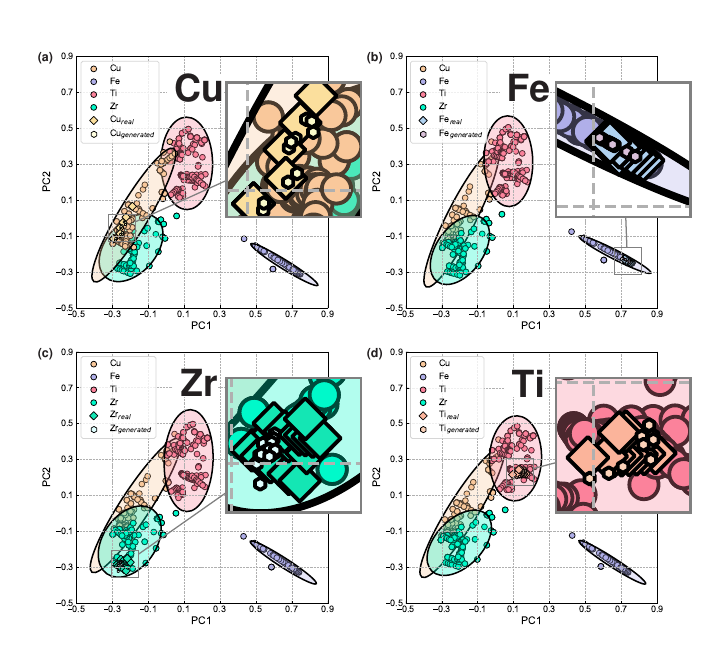}
	\end{center}
    \vspace{-35pt}
 	\caption{Inverse alloy design using conditional GAN. Two-dimensional PCA plots compare the generated samples from AlloyGAN with the original ones across four alloy categories. Data points in the latent space are color-coded to represent their respective categories. Confidence ellipses in corresponding colors highlight the distribution of each category, showing the alignment between generated and original data.}
	\label{fig4}
\end{figure*}

\begin{figure*}[ht]
	\begin{center}
		\centering
        \vspace{-25pt}
		\includegraphics[width=1\textwidth]{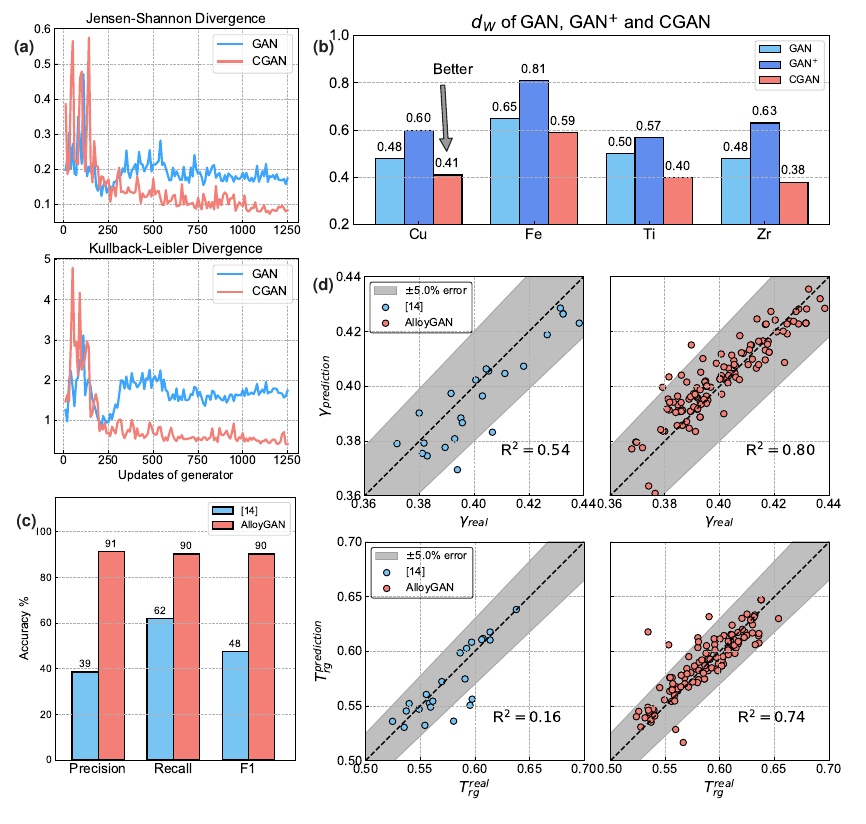}
	\end{center}
\vspace{-25pt}	\caption{Performance evaluation of machine-learning models. (a) Training stability comparison between GAN and CGAN models over 1250 generator updates, assessed using Jensen-Shannon (JS) divergence and Kullback-Leibler (KL) divergence. (b) Wasserstein distance ($d_w$) comparison among GAN, GAN+, and CGAN models. GAN uses only alloy composition as input, while GAN+ incorporates both composition and property features as inputs. (c) Classification accuracy comparison between our workflow and prior approaches, showing a significant improvement in categorizing alloys based on specific criteria. (d) Regression accuracy comparison for the properties $T_{rg}$ and $\gamma$.}
	\label{complex}
\end{figure*}

To improve the performance of inverse alloy design, we developed a CGAN model that incorporates conditional information. Unlike the traditional GAN, which generates alloy compositions randomly, CGAN integrates specific conditional inputs into the latent space, enabling precise control over the generated outputs. This capability allows CGAN to design alloys with targeted properties, offering a significant advantage in materials design. For model evaluation, 10\% of each alloy dataset was reserved as a test set. As shown in Fig.~\ref{fig4}, the generated alloys from CGAN align well within the distribution range of data from the literature. Notably, even in cases where significant overlaps exist in the latent space, such as between Cu and Zr, CGAN demonstrates robust generalization capabilities.

To quantitatively assess the quality of the generated alloys, we calculated the Kullback-Leibler (KL)~\cite{kullback1951information} and Jensen-Shannon (JS)~\cite{tishby2000information} divergences to measure the difference between the real and synthetic distributions, where lower values of these metrics indicate a closer match. Specifically, the KL divergence quantifies the information loss when one distribution approximates another, while the JS divergence offers a symmetric and bounded measure of similarity between two distributions. Both metrics are commonly used to evaluate the generative model quality. The formulas for the KL and JS divergences are Supplementary Eqs.~\ref{eq:kl}, \ref{eq:js}.

Fig.~\ref{complex}(a) compares the training dynamics of traditional GAN and CGAN on the Ti dataset. CGAN achieves more stable training, with KL divergence consistently $<$1 and JS divergence $<$0.1 throughout the process. In contrast to the traditional GAN models, CGAN generates alloys that align more closely to the real distribution, demonstrating superior stability and performance in inverse alloy design.

The superior performance of CGANs compared to GANs can be attributed to its integration of thermodynamic quantities as conditional inputs during training. These thermodynamic features enhance the generation process without overfitting the training data, which improves model effectiveness. To investigate whether directly using thermodynamic data for training would yield similar improvements, we introduced a modified GAN model (denoted as GAN\textsuperscript{+}) that incorporates thermodynamic data as direct input rather than auxiliary conditional information.

To evaluate the relative performance of GAN, GAN\textsuperscript{+}, and CGAN, we utilized the Wasserstein distance ($d_W$)~\cite{vaserstein1969markov} as a metric. The Wasserstein distance quantifies the difference between two probability distributions, with smaller values indicating greater similarity between generated and real data distributions. This metric is particularly useful for assessing how well generative models capture the underlying data distribution. The formulas for calculating Wasserstein distance are Supplementary Eq.~\ref{eq:wd}.

In the Cu dataset, when the GAN model was trained using only alloy compositions, the Wasserstein distance between the generated and original data was 0.48 (Fig.~\ref{complex}(b)). Surprisingly, incorporating thermodynamic data directly into the GAN training process (GAN\textsuperscript{+}) increased the Wasserstein distance to 0.60, signifying worse performance. This result underscores a limitation of the traditional GAN approach: adding more features directly to the input does not necessarily enhance model performance. Instead, it divides the model's focus, making it challenging to distinguish and learn the relationships between alloy compositions and properties effectively. The issue becomes more pronounced as additional features are included.

In contrast, CGAN achieves a Wasserstein distance of 0.41 for the Cu dataset, demonstrating a significant improvement in the ability to generate alloy compositions that closely align with the original data distribution. This enhancement highlights the effectiveness of incorporating thermodynamic features as conditional inputs rather than direct inputs. By integrating these features conditionally, CGAN preserves the integrity of alloy composition data while simultaneously leveraging the relationship between composition and properties.

Overall, CGANs address the limitations of GANs, generating more precise and reliable alloy compositions. Using thermodynamic properties as conditional inputs strengthens the inverse design process, producing alloy samples that are consistent with real-world data and align with targeted property requirements. This robust framework positions CGAN as a powerful tool for advancing the inverse design of alloys with tailored properties.

\subsection{Downstream Tasks}

We extend the application of our AlloyGAN framework to a range of downstream classification and regression tasks to validate its efficiency, adaptability, and versatility. These tasks demonstrate the framework's ability to generalize across diverse datasets and applications, highlighting its potential as a robust tool for addressing complex challenges in materials design. 
\subsubsection{Classification task}

GFA classification has been a focus of ML applications in materials science. However, many existing ML models rely heavily on manually calculated or experimentally derived material features, limiting their scalability and usability. These approaches often require recalculating features for new datasets, significantly complicating the training process.

To address these challenges, we constructed a classification model~\cite{lecun2015deep} using our AlloyGAN-generated data, leveraging only alloy compositions as input features to predict GFA. This simplifies the training process by removing the need for extensive feature engineering or experimental data, making the model more efficient, user-friendly, and scalable.

Using the AlloyGAN-enhanced dataset, the classification model effectively learns and predicts GFA with high accuracy. Alloys are categorized into two groups based on their reduced glass transition temperature $T_{rg}$~\cite{lu2000correlation}: alloys with good GFA ($T_{rg}>0.6$) and poor GFA ($T_{rg}<0.6$). The dataset, comprising 1,338 alloy compositions collected through LLM-assisted data mining, was split into 80\% training and 20\% test sets. We use precision, recall and $F^1_\text{score}$ to measure the performance of models. Detailed formulas are shown in Supplementary Notes 1.2.

Precision measures the proportion of correctly predicted positives among all positive predictions, while recall assesses the proportion of true positives identified. The $F^1_\text{score}$, as the harmonic mean of precision and recall, balances these metrics to provide a comprehensive evaluation. As shown in Fig.~\ref{complex}(c), our approach achieved precision, recall, and  $F^1_\text{score}$ values consistently around 0.90, demonstrating exceptional accuracy. This underscores the effectiveness of AlloyGAN in streamlining classification tasks and advancing materials science by providing a scalable, high-accuracy solution for alloy design and analysis.

\subsubsection{Regression task}

We conducted regression tests to further validate the effectiveness of the AlloyGAN framework in downstream tasks. The training and test sets are identical to those in the classification tasks. In Fig.~\ref{complex}(d), we compare the true and predicted values of $T_{rg}$~\cite{lu2000correlation} and GFA criteria $\gamma$~\cite{lu2002new} on the test set. The results highlight the superior predictive performance of our method compared to previous approaches.

We summarize the performance metrics for our framework and earlier methods across the regression tasks (details in Supplementary Table~S2). Notably, our model achieved $R^2$ values of 0.74 for $T_{rg}$ and 0.80 for $\gamma$, significantly outperforming the $R^2$ values of 0.16 and 0.54 for previous methods. These improvements underscore the capability of AlloyGAN to achieve higher accuracy and robustness in predicting target properties.

The results demonstrate that AlloyGAN generates high-quality alloy compositions and excels at uncovering and modeling the intrinsic relationships between alloy compositions and their properties. This multitasking ability enhances both the interpretability and predictive power of the model, showcasing its strong generalization performance and potential for advancing alloy design through regression tasks.

\subsection{Experimental Validation}

To assess the reliability of our approach, we generated metallic glass candidates using AlloyGAN, conditioned on thermodynamic parameters--specifically, the reduced glass transition temperature $T_{rg}$~\cite{lu2000correlation} and $\gamma$~\cite{lu2002new}. The model was designed to generate compositions where $T_{rg}\geq0.6$ or $\gamma\geq0.4$, indicating strong GFA.
This targeted generation strategy enabled the identification of compositions with enhanced thermal stability, making them promising candidates for experimental validation (listed in Supplementary Table~\ref{tab:alloy_data}).

\begin{figure*}[ht]
	\begin{center}
		\centering
        \vspace{-10pt}
		\includegraphics[width=1\textwidth]{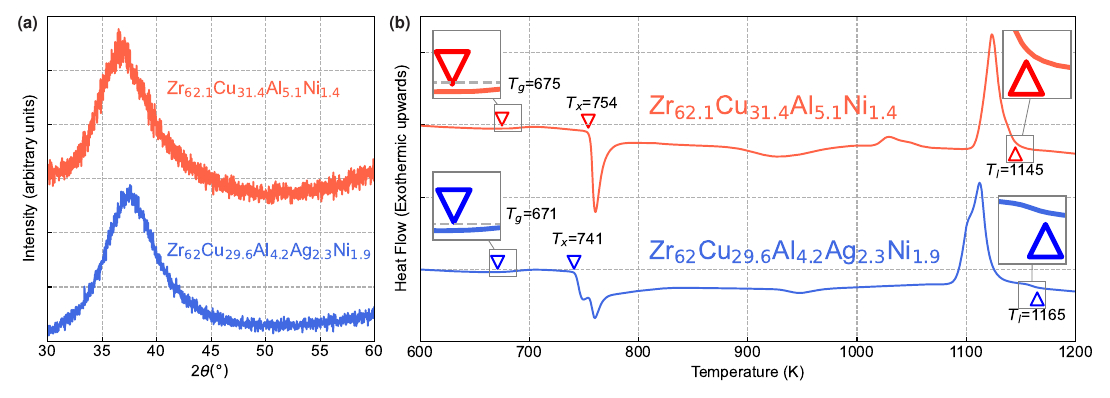}
	\end{center}
\vspace{-10pt}	\caption{Experiment validation of two metallic glass candidates, {Zr$_{62.1}$Cu$_{31.4}$Al$_{5.1}$Ni$_{1.4}$} and {Zr$_{62}$Cu$_{29.6}$Al$_{4.2}$Ag$_{2.3}$Ni$_{1.9}$}. (a) X-ray diffraction (XRD) patterns of 5mm cast rods, representing amorphous structures. (b) Differential scanning calorimetry (DSC) curves showing the measured $T_{g}$,  $T_{x}$ and $T_l$. }
	\label{fig:exp}
\end{figure*}

For experimental validation, we selected two Zr-based alloy compositions, {Zr$_{62.1}$Cu$_{31.4}$Al$_{5.1}$Ni$_{1.4}$} and {Zr$_{62}$Cu$_{29.6}$Al$_{4.2}$Ag$_{2.3}$Ni$_{1.9}$}, based on their high $T_{rg}$ and $\gamma$, as well as the cost consideration associated with Ag~\cite{cost_report}. 5-mm-diameter rod samples were synthesized, and X-ray diffraction (XRD) analysis~\cite{book_xrd} confirmed their amorphous nature~(Fig.~\ref{fig:exp}(a)).
To further characterize their thermal properties, differential scanning calorimetry (DSC)~\cite{book_dsc,greer2015metallic} was performed to measure $T_{g}$, $T_{x}$ and $T_{l}$ (Fig.~\ref{fig:exp}(b); see Methods for details).

As summarized in Table~\ref{tab:exp_results}, the experimentally measured thermodynamic properties align well with the predicted values by CGAN. For {Zr$_{62.1}$Cu$_{31.4}$Al$_{5.1}$Ni$_{1.4}$}, the model's predictions for $T_{g}$, $T_{x}$, $T_{l}$, $T_{rg}$, and $\gamma$ show a maximum deviation of 5.8\% from experimental results. 
Similarly, for {Zr$_{62}$Cu$_{29.6}$Al$_{4.2}$Ag$_{2.3}$Ni$_{1.9}$}, the maximum deviation is 7.7\%. Despite minor discrepancies, the results highlight the predictive capability of CGAN in estimating thermodynamic properties of novel alloys, demonstrating its potential as an effective tool for guiding materials discovery.

\begin{table}[ht]
\caption{Experimental validation of metallic glass candidates generated by the AlloyGAN framework. Comparison of predicted and experimentally measured thermodynamic properties.}
\label{tab:exp_results}
\begin{tabular}{cccccc}
\toprule
\multicolumn{6}{c}{Zr$_{62.1}$Cu$_{31.4}$Al$_{5.1}$Ni$_{1.4}$}   \\
\midrule
Property  & $T_g$  (K) & $T_x$(K) & $T_l$  (K)& $T_{rg}$ & $\gamma$ \\
\midrule Prediction & 699 & 769 & 1121 & 0.623 & 0.422 \\
 Experiment & 675 & 754 & 1145 &  0.590&  0.414\\
 Bias &24   & 15  & 24 & 0.034 &0.008 \\
 Relative Error (\%) & \textbf{3.6}  & \textbf{2.0}  &\textbf{2.1} &\textbf{5.8}&\textbf{2.0} \\
\toprule
\multicolumn{6}{c}{Zr$_{62}$Cu$_{29.6}$Al$_{4.2}$Ag$_{2.3}$Ni$_{1.9}$}   \\
 \midrule
Property  & $T_g$  (K) & $T_x$(K) & $T_l$  (K)& $T_{rg}$ & $\gamma$ \\
\midrule Prediction & 698 & 765 & 1125 & 0.620 & 0.420 \\
 Experiment & 671& 741 & 1165& 0.576& 0.404
\\
 Bias &27  & 24 &40 &0.044 &0.016\\
 Relative Error (\%) & \textbf{4.0}  & \textbf{3.2} &\textbf{3.4 }&\textbf{7.7}&\textbf{4.0} \\
\toprule
\end{tabular}
\end{table}

\section{Discussion}

In this work, we introduce AlloyGAN, a CGAN-based framework assisted by LLM for inverse materials design, using alloy composition optimization as a case study. By integrating generative AI, LLM-driven data mining, and experimental validation, AlloyGAN addresses key challenges in data sparsity, efficiency, and accuracy that have hindered the practical application of generative models in materials discovery. 
Unlike conventional generative AI approaches, AlloyGAN incorporates thermodynamic properties as conditional inputs, enabling more accurate and interpretable predictions across diverse materials systems. 
The framework achieves thermophysical property predictions within 8\% deviation from experimental results, leveraging a training dataset more than six times larger than those used in prior studies~\cite{zhou2023generative}, thanks to automated literature extraction via LLMs.

Despite its strong predictive performance, several challenges remain in further enhancing AlloyGAN for broader materials discovery. First, expanding the training dataset is critical, as high-quality experimental data remains sparse for many material systems. 
While AlloyGAN leverages LLM-assisted data mining to enhance data diversity, future work should integrate multi-modal datasets, including high-throughput simulations and experimental measurements, to further improve model accuracy. 
Second, refining conditional constraints could enhance its applicability beyond thermodynamic properties. Incorporating additional descriptors--such as mechanical and electronic properties--may enable more comprehensive inverse design strategies. 
Finally, generalization beyond alloys remains an open challenge. While AlloyGAN has demonstrated success in alloy design, extending this framework to other material classes--such as battery materials, superconductors, and functional ceramics--requires further validation and adaptation.

Furthermore, the integration of LLMs into materials informatics highlights the transformative potential of natural language processing in materials science~\cite{liu2024,hu2024}.
By automating literature mining and descriptor extraction, LLMs bridge the gap between experimental and computational approaches, fostering a data-driven and scalable paradigm for materials research. 
This synergy between AI-driven design, high-throughput computation, experimental validation, and the emerging reasoning ability of LLMs to think and act like materials scientists pave the way for a new era of autonomous materials discovery, offering new opportunities for the rapid and efficient design of next-generation materials.

\section{Methodology}
\subsection{Dataset}
 
The lack of a unified, publicly available dataset for multi-component alloys~\cite{raabe2023accelerating} presents a significant challenge for data-driven alloy design and discovery. 
Existing alloy datasets are often fragmented, manually extracted from literature, and constrained by limited sample sizes, restricting the effectiveness of machine learning models. 
For example, previous research~\cite{zhou2023generative} relied on datasets with fewer than 300 entries, leading to suboptimal model performance. 
To address these limitations, we developed an integrated framework for automated alloy data mining and inverse alloy design. 
This framework systematically extracts alloy data from the literature, incorporates experimental validation, and continuously updates the dataset with newly generated alloy compositions, enabling a self-improving cycle. Additionally, the dataset serves as a foundation for downstream tasks such as classification and regression, further enhancing its utility.

Traditional methods for acquiring alloy data involve labor-intensive experimental synthesis and characterization or manual literature extraction.
To streamline this process, we employed a LLM-driven approach for automated data extraction. 
Specifically, we utilized Uni-SMART~\cite{cai2024uni}, a state-of-the-art LLM capable of processing diverse scientific modalities, including textual descriptions, diagrams, mathematical equations, molecular structures, and chemical reaction equations. 
Its multimodal capability ensures a more comprehensive and accurate extraction of scientific data, significantly reducing human effort.

\begin{figure}[h]
	\begin{center}
		\centering
        \includegraphics[width=0.52\textwidth]{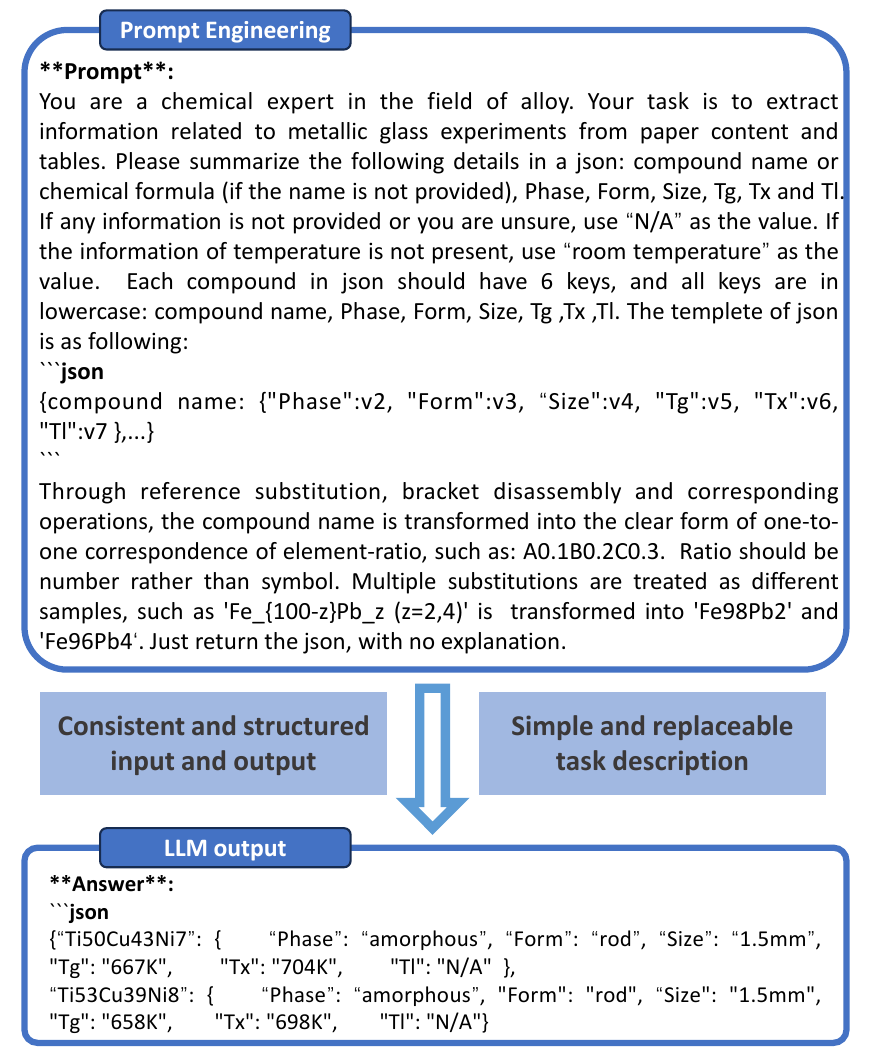}
	\end{center}
	\caption{Example prompt used for extracting alloy information from literature using Uni-SMART~\cite{cai2024uni}.}
	\label{fig:prompt}
\end{figure}

To ensure data consistency and reliability, we designed structured prompts tailored to guide Uni-SMART in identifying and organizing alloy-related information from literature. 
A `few-shot' learning approach was implemented, embedding representative examples within the prompts to specify the desired content, format, and processing details. 
The extracted data were then returned in a standardized JSON format, facilitating seamless integration into our alloy design pipeline. An example of a structured prompt is shown in Fig.~\ref{fig:prompt}.

By leveraging Uni-SMART LLM, we systematically mined alloy compositions and thermodynamic properties from over 200 published sources, resulting in a dataset of 1,338 entries--a substantial expansion of the existing data pool. This scalable and automated data-mining approach provides a foundation for accelerating inverse alloy design and advancing materials informatics applications.

\subsection{Descriptor construction}

Feature engineering plays a critical role in machine learning models, as the choice and quality of descriptors directly influence predictive performance. 
To optimize model performance, we systematically constructed and selected key features from the extracted literature data, ensuring the inclusion of informative and representative descriptors.

Building upon composition-based descriptors proposed in~\cite{zhou2023generative}, we incorporated thermodynamic properties to enhance the predictive capability of our framework. 
Specifically, we extracted $T_{x}$, $T_{g}$ and $T_{l}$ from the literature as primary thermodynamic descriptors. 
Additionally, we derived several GFA parameters as auxiliary features, all expressed as mathematical functions of $T_{x}$, $T_{g}$ and $T_{l}$. This automated descriptor generation process ensures consistency and scalability in data preprocessing.

A comprehensive summary of the constructed descriptors is provided in Table~S1. The final dataset comprises 1,338 alloy entries, each containing a composition descriptor alongside 26 thermodynamic features. 
This enriched dataset significantly improves the representation of alloy systems, facilitating robust model training and enhancing the predictive accuracy of our ML framework.

\subsection{Model Construction}

We developed a generative deep learning framework for inverse alloy design based on a CGAN, enabling the targeted generation of alloy compositions with desired properties. 
GANs~\cite{goodfellow2014generative} generate samples by mapping a latent noise distribution to an alloy composition space, ensuring that the generated compositions follow the statistical distribution of the training data. 
However, standard GANs lack the ability to design alloy compositions under predefined property constraints, limiting their applicability in inverse materials design. 
To overcome this challenge, we extended the GAN architecture by introducing conditional constraints, allowing the model to generate compositions that satisfy specific design requirements (Fig.~S1).

The CGAN framework consists of two key components:

1. \textbf{Generator G:}
The generator takes a noise vector $z$ sampled from a prior distribution $P_z$, along with a conditional input $c$, which encodes the target alloy properties or design constraints. 
By learning the underlying relationship between alloy compositions and their associated properties, the generator synthesizes new compositions that are both statistically realistic and align with the specified property constraints. 
This conditional mechanism enables targeted exploration of the compositional space, allowing for property-driven alloy generation.

2. \textbf{Discriminator D:}
The discriminator is trained to distinguish between real alloy compositions, sampled from the training data distribution $P_{data}(X|c)$ and synthetic compositions generated by $G$, which follow the synthetic distribution $P_g(X|c)$. 
Both real and generated compositions are presented to the discriminator along with the corresponding conditional input $c$, enabling it to assess whether a given alloy composition aligns with the specified property constraints.

During training, the generator iteratively refines its outputs to better match the real data distribution while satisfying the given design conditions. 
This is achieved through an adversarial process, where $G$ continuously improves its ability to generate realistic compositions, while $D$ improves its ability to distinguish between real and synthetic alloys. 
Over successive training iterations, $G$ learns to capture the conditional distribution $P_{data}(X|c)$, ultimately generating compositions that exhibit both the statistical patterns of real data and the desired property constraints with high accuracy.

By integrating property-driven constraints, our CGAN-based framework significantly enhances the accuracy and applicability of generative AI for alloy design. 
The model enables researchers to generate new alloy compositions that meet specific performance criteria, bridging the gap between data-driven generative modeling and practical materials design. 
This approach offers a scalable and efficient pathway for navigating complex compositional spaces, accelerating the discovery of novel alloys with tailored properties.

\subsection{Experimental Methods}
Master alloy ingots were prepared by arc-melting ultrasonically cleaned pure metals Zr, Cu, Ni, Al, Ag in Ar atmosphere. 
The purity of the metals is higher than 99.9\%. 
To ensure chemical homogeneity, the ingots were remelted at least four times before being used for further processing. 
Rod samples were fabricated by suction-casting the remelted ingots into a copper mold using the arc-melting technique.

The phase formation of the as-cast rod samples was analyzed using XRD on a Rigaku SmartLab diffractometer with Cu-K$\alpha$ radiation. The thermal properties of the samples were examined using a Simultaneous Thermal Analyzer (PerkinElmer STA8000) at a heating rate of 20 K/min.

\section{Data Availability}

The source data and code for AlloyGAN are available in the Github repository: \href{https://github.com/photon-git/AlloyGAN}{https://github.com/photon-git/AlloyGAN}. 
\section{Acknowledgments}
Y.H. and Q.W. were supported by the National Natural Science Foundation of China (NSFC) 62272172.
W.L, Z.L, and Y.L were supported by the NSFC 52331007.
T.W. acknowledges support by The University of Hong Kong via seed fund 2409100597.

\def\bibsection{\section*{\refname}}

\bibliography{refs.bib}

\begin{thebibliography}{51}%
\makeatletter
\providecommand \@ifxundefined [1]{%
 \@ifx{#1\undefined}
}%
\providecommand \@ifnum [1]{%
 \ifnum #1\expandafter \@firstoftwo
 \else \expandafter \@secondoftwo
 \fi
}%
\providecommand \@ifx [1]{%
 \ifx #1\expandafter \@firstoftwo
 \else \expandafter \@secondoftwo
 \fi
}%
\providecommand \natexlab [1]{#1}%
\providecommand \enquote  [1]{``#1''}%
\providecommand \bibnamefont  [1]{#1}%
\providecommand \bibfnamefont [1]{#1}%
\providecommand \citenamefont [1]{#1}%
\providecommand \href@noop [0]{\@secondoftwo}%
\providecommand \href [0]{\begingroup \@sanitize@url \@href}%
\providecommand \@href[1]{\@@startlink{#1}\@@href}%
\providecommand \@@href[1]{\endgroup#1\@@endlink}%
\providecommand \@sanitize@url [0]{\catcode `\\12\catcode `\$12\catcode
  `\&12\catcode `\#12\catcode `\^12\catcode `\_12\catcode `\%12\relax}%
\providecommand \@@startlink[1]{}%
\providecommand \@@endlink[0]{}%
\providecommand \url  [0]{\begingroup\@sanitize@url \@url }%
\providecommand \@url [1]{\endgroup\@href {#1}{\urlprefix }}%
\providecommand \urlprefix  [0]{URL }%
\providecommand \Eprint [0]{\href }%
\providecommand \doibase [0]{https://doi.org/}%
\providecommand \selectlanguage [0]{\@gobble}%
\providecommand \bibinfo  [0]{\@secondoftwo}%
\providecommand \bibfield  [0]{\@secondoftwo}%
\providecommand \translation [1]{[#1]}%
\providecommand \BibitemOpen [0]{}%
\providecommand \bibitemStop [0]{}%
\providecommand \bibitemNoStop [0]{.\EOS\space}%
\providecommand \EOS [0]{\spacefactor3000\relax}%
\providecommand \BibitemShut  [1]{\csname bibitem#1\endcsname}%
\let\auto@bib@innerbib\@empty
\bibitem [{\citenamefont {Raabe}\ \emph {et~al.}(2023)\citenamefont {Raabe},
  \citenamefont {Mianroodi},\ and\ \citenamefont
  {Neugebauer}}]{raabe2023accelerating}%
  \BibitemOpen
  \bibfield  {author} {\bibinfo {author} {\bibfnamefont {D.}~\bibnamefont
  {Raabe}}, \bibinfo {author} {\bibfnamefont {J.~R.}\ \bibnamefont
  {Mianroodi}},\ and\ \bibinfo {author} {\bibfnamefont {J.}~\bibnamefont
  {Neugebauer}},\ }\bibfield  {title} {\bibinfo {title} {Accelerating the
  design of compositionally complex materials via physics-informed artificial
  intelligence},\ }\href@noop {} {\bibfield  {journal} {\bibinfo  {journal}
  {Nature Computational Science}\ }\textbf {\bibinfo {volume} {3}},\ \bibinfo
  {pages} {198} (\bibinfo {year} {2023})}\BibitemShut {NoStop}%
\bibitem [{\citenamefont {Juan}\ \emph {et~al.}(2021)\citenamefont {Juan},
  \citenamefont {Dai}, \citenamefont {Yang},\ and\ \citenamefont
  {Zhang}}]{juan2021accelerating}%
  \BibitemOpen
  \bibfield  {author} {\bibinfo {author} {\bibfnamefont {Y.}~\bibnamefont
  {Juan}}, \bibinfo {author} {\bibfnamefont {Y.}~\bibnamefont {Dai}}, \bibinfo
  {author} {\bibfnamefont {Y.}~\bibnamefont {Yang}},\ and\ \bibinfo {author}
  {\bibfnamefont {J.}~\bibnamefont {Zhang}},\ }\bibfield  {title} {\bibinfo
  {title} {Accelerating materials discovery using machine learning},\
  }\href@noop {} {\bibfield  {journal} {\bibinfo  {journal} {Journal of
  Materials Science \& Technology}\ }\textbf {\bibinfo {volume} {79}},\
  \bibinfo {pages} {178} (\bibinfo {year} {2021})}\BibitemShut {NoStop}%
\bibitem [{\citenamefont {Li}\ \emph {et~al.}(2020)\citenamefont {Li},
  \citenamefont {Lim}, \citenamefont {Yang}, \citenamefont {Ren}, \citenamefont
  {Raghavan}, \citenamefont {Chen}, \citenamefont {Buonassisi},\ and\
  \citenamefont {Wang}}]{li2020ai}%
  \BibitemOpen
  \bibfield  {author} {\bibinfo {author} {\bibfnamefont {J.}~\bibnamefont
  {Li}}, \bibinfo {author} {\bibfnamefont {K.}~\bibnamefont {Lim}}, \bibinfo
  {author} {\bibfnamefont {H.}~\bibnamefont {Yang}}, \bibinfo {author}
  {\bibfnamefont {Z.}~\bibnamefont {Ren}}, \bibinfo {author} {\bibfnamefont
  {S.}~\bibnamefont {Raghavan}}, \bibinfo {author} {\bibfnamefont {P.-Y.}\
  \bibnamefont {Chen}}, \bibinfo {author} {\bibfnamefont {T.}~\bibnamefont
  {Buonassisi}},\ and\ \bibinfo {author} {\bibfnamefont {X.}~\bibnamefont
  {Wang}},\ }\bibfield  {title} {\bibinfo {title} {Ai applications through the
  whole life cycle of material discovery},\ }\href@noop {} {\bibfield
  {journal} {\bibinfo  {journal} {Matter}\ }\textbf {\bibinfo {volume} {3}},\
  \bibinfo {pages} {393} (\bibinfo {year} {2020})}\BibitemShut {NoStop}%
\bibitem [{\citenamefont {Louie}\ \emph {et~al.}(2021)\citenamefont {Louie},
  \citenamefont {Chan}, \citenamefont {da~Jornada}, \citenamefont {Li},\ and\
  \citenamefont {Qiu}}]{louie2021discovering}%
  \BibitemOpen
  \bibfield  {author} {\bibinfo {author} {\bibfnamefont {S.~G.}\ \bibnamefont
  {Louie}}, \bibinfo {author} {\bibfnamefont {Y.-H.}\ \bibnamefont {Chan}},
  \bibinfo {author} {\bibfnamefont {F.~H.}\ \bibnamefont {da~Jornada}},
  \bibinfo {author} {\bibfnamefont {Z.}~\bibnamefont {Li}},\ and\ \bibinfo
  {author} {\bibfnamefont {D.~Y.}\ \bibnamefont {Qiu}},\ }\bibfield  {title}
  {\bibinfo {title} {Discovering and understanding materials through
  computation},\ }\href@noop {} {\bibfield  {journal} {\bibinfo  {journal}
  {Nature Materials}\ }\textbf {\bibinfo {volume} {20}},\ \bibinfo {pages}
  {728} (\bibinfo {year} {2021})}\BibitemShut {NoStop}%
\bibitem [{\citenamefont {Merchant}\ \emph {et~al.}(2023)\citenamefont
  {Merchant}, \citenamefont {Batzner}, \citenamefont {Schoenholz},
  \citenamefont {Aykol}, \citenamefont {Cheon},\ and\ \citenamefont
  {Cubuk}}]{merchant2023scaling}%
  \BibitemOpen
  \bibfield  {author} {\bibinfo {author} {\bibfnamefont {A.}~\bibnamefont
  {Merchant}}, \bibinfo {author} {\bibfnamefont {S.}~\bibnamefont {Batzner}},
  \bibinfo {author} {\bibfnamefont {S.~S.}\ \bibnamefont {Schoenholz}},
  \bibinfo {author} {\bibfnamefont {M.}~\bibnamefont {Aykol}}, \bibinfo
  {author} {\bibfnamefont {G.}~\bibnamefont {Cheon}},\ and\ \bibinfo {author}
  {\bibfnamefont {E.~D.}\ \bibnamefont {Cubuk}},\ }\bibfield  {title} {\bibinfo
  {title} {Scaling deep learning for materials discovery},\ }\href@noop {}
  {\bibfield  {journal} {\bibinfo  {journal} {Nature}\ }\textbf {\bibinfo
  {volume} {624}},\ \bibinfo {pages} {80} (\bibinfo {year} {2023})}\BibitemShut
  {NoStop}%
\bibitem [{\citenamefont {Liu}\ \emph {et~al.}(2020)\citenamefont {Liu},
  \citenamefont {Li}, \citenamefont {He}, \citenamefont {Liang}, \citenamefont
  {Zhou}, \citenamefont {Ma}, \citenamefont {Yang},\ and\ \citenamefont
  {Shen}}]{liu2020machine}%
  \BibitemOpen
  \bibfield  {author} {\bibinfo {author} {\bibfnamefont {X.}~\bibnamefont
  {Liu}}, \bibinfo {author} {\bibfnamefont {X.}~\bibnamefont {Li}}, \bibinfo
  {author} {\bibfnamefont {Q.}~\bibnamefont {He}}, \bibinfo {author}
  {\bibfnamefont {D.}~\bibnamefont {Liang}}, \bibinfo {author} {\bibfnamefont
  {Z.}~\bibnamefont {Zhou}}, \bibinfo {author} {\bibfnamefont {J.}~\bibnamefont
  {Ma}}, \bibinfo {author} {\bibfnamefont {Y.}~\bibnamefont {Yang}},\ and\
  \bibinfo {author} {\bibfnamefont {J.}~\bibnamefont {Shen}},\ }\bibfield
  {title} {\bibinfo {title} {Machine learning-based glass formation prediction
  in multicomponent alloys},\ }\href@noop {} {\bibfield  {journal} {\bibinfo
  {journal} {Acta Materialia}\ }\textbf {\bibinfo {volume} {201}},\ \bibinfo
  {pages} {182} (\bibinfo {year} {2020})}\BibitemShut {NoStop}%
\bibitem [{\citenamefont {Ward}\ \emph {et~al.}(2016)\citenamefont {Ward},
  \citenamefont {Agrawal}, \citenamefont {Choudhary},\ and\ \citenamefont
  {Wolverton}}]{ward2016general}%
  \BibitemOpen
  \bibfield  {author} {\bibinfo {author} {\bibfnamefont {L.}~\bibnamefont
  {Ward}}, \bibinfo {author} {\bibfnamefont {A.}~\bibnamefont {Agrawal}},
  \bibinfo {author} {\bibfnamefont {A.}~\bibnamefont {Choudhary}},\ and\
  \bibinfo {author} {\bibfnamefont {C.}~\bibnamefont {Wolverton}},\ }\bibfield
  {title} {\bibinfo {title} {A general-purpose machine learning framework for
  predicting properties of inorganic materials},\ }\href@noop {} {\bibfield
  {journal} {\bibinfo  {journal} {npj Computational Materials}\ }\textbf
  {\bibinfo {volume} {2}},\ \bibinfo {pages} {1} (\bibinfo {year}
  {2016})}\BibitemShut {NoStop}%
\bibitem [{\citenamefont {Zhou}\ \emph {et~al.}(2021)\citenamefont {Zhou},
  \citenamefont {He}, \citenamefont {Liu}, \citenamefont {Wang}, \citenamefont
  {Luan}, \citenamefont {Liu},\ and\ \citenamefont {Yang}}]{zhou2021rational}%
  \BibitemOpen
  \bibfield  {author} {\bibinfo {author} {\bibfnamefont {Z.}~\bibnamefont
  {Zhou}}, \bibinfo {author} {\bibfnamefont {Q.}~\bibnamefont {He}}, \bibinfo
  {author} {\bibfnamefont {X.}~\bibnamefont {Liu}}, \bibinfo {author}
  {\bibfnamefont {Q.}~\bibnamefont {Wang}}, \bibinfo {author} {\bibfnamefont
  {J.}~\bibnamefont {Luan}}, \bibinfo {author} {\bibfnamefont {C.}~\bibnamefont
  {Liu}},\ and\ \bibinfo {author} {\bibfnamefont {Y.}~\bibnamefont {Yang}},\
  }\bibfield  {title} {\bibinfo {title} {Rational design of chemically complex
  metallic glasses by hybrid modeling guided machine learning},\ }\href@noop {}
  {\bibfield  {journal} {\bibinfo  {journal} {npj Computational Materials}\
  }\textbf {\bibinfo {volume} {7}},\ \bibinfo {pages} {138} (\bibinfo {year}
  {2021})}\BibitemShut {NoStop}%
\bibitem [{\citenamefont {Ward}\ \emph {et~al.}(2018)\citenamefont {Ward},
  \citenamefont {O'Keeffe}, \citenamefont {Stevick}, \citenamefont {Jelbert},
  \citenamefont {Aykol},\ and\ \citenamefont {Wolverton}}]{ward2018machine}%
  \BibitemOpen
  \bibfield  {author} {\bibinfo {author} {\bibfnamefont {L.}~\bibnamefont
  {Ward}}, \bibinfo {author} {\bibfnamefont {S.~C.}\ \bibnamefont {O'Keeffe}},
  \bibinfo {author} {\bibfnamefont {J.}~\bibnamefont {Stevick}}, \bibinfo
  {author} {\bibfnamefont {G.~R.}\ \bibnamefont {Jelbert}}, \bibinfo {author}
  {\bibfnamefont {M.}~\bibnamefont {Aykol}},\ and\ \bibinfo {author}
  {\bibfnamefont {C.}~\bibnamefont {Wolverton}},\ }\bibfield  {title} {\bibinfo
  {title} {A machine learning approach for engineering bulk metallic glass
  alloys},\ }\href@noop {} {\bibfield  {journal} {\bibinfo  {journal} {Acta
  Materialia}\ }\textbf {\bibinfo {volume} {159}},\ \bibinfo {pages} {102}
  (\bibinfo {year} {2018})}\BibitemShut {NoStop}%
\bibitem [{\citenamefont {Xiong}\ \emph {et~al.}(2019)\citenamefont {Xiong},
  \citenamefont {Zhang},\ and\ \citenamefont {Shi}}]{xiong2019machine}%
  \BibitemOpen
  \bibfield  {author} {\bibinfo {author} {\bibfnamefont {J.}~\bibnamefont
  {Xiong}}, \bibinfo {author} {\bibfnamefont {T.-Y.}\ \bibnamefont {Zhang}},\
  and\ \bibinfo {author} {\bibfnamefont {S.-Q.}\ \bibnamefont {Shi}},\
  }\bibfield  {title} {\bibinfo {title} {Machine learning prediction of elastic
  properties and glass-forming ability of bulk metallic glasses},\ }\href@noop
  {} {\bibfield  {journal} {\bibinfo  {journal} {MRS Communications}\ }\textbf
  {\bibinfo {volume} {9}},\ \bibinfo {pages} {576} (\bibinfo {year}
  {2019})}\BibitemShut {NoStop}%
\bibitem [{\citenamefont {Xiong}\ \emph {et~al.}(2020)\citenamefont {Xiong},
  \citenamefont {Shi},\ and\ \citenamefont {Zhang}}]{xiong2020machine}%
  \BibitemOpen
  \bibfield  {author} {\bibinfo {author} {\bibfnamefont {J.}~\bibnamefont
  {Xiong}}, \bibinfo {author} {\bibfnamefont {S.-Q.}\ \bibnamefont {Shi}},\
  and\ \bibinfo {author} {\bibfnamefont {T.-Y.}\ \bibnamefont {Zhang}},\
  }\bibfield  {title} {\bibinfo {title} {A machine-learning approach to
  predicting and understanding the properties of amorphous metallic alloys},\
  }\href@noop {} {\bibfield  {journal} {\bibinfo  {journal} {Materials \&
  design}\ }\textbf {\bibinfo {volume} {187}},\ \bibinfo {pages} {108378}
  (\bibinfo {year} {2020})}\BibitemShut {NoStop}%
\bibitem [{\citenamefont {Wang}(2014)}]{wang2014high}%
  \BibitemOpen
  \bibfield  {author} {\bibinfo {author} {\bibfnamefont {W.}~\bibnamefont
  {Wang}},\ }\bibfield  {title} {\bibinfo {title} {High-entropy metallic
  glasses},\ }\href@noop {} {\bibfield  {journal} {\bibinfo  {journal} {Jom}\
  }\textbf {\bibinfo {volume} {66}},\ \bibinfo {pages} {2067} (\bibinfo {year}
  {2014})}\BibitemShut {NoStop}%
\bibitem [{\citenamefont {Zhang}\ \emph {et~al.}(2022)\citenamefont {Zhang},
  \citenamefont {Zhou}, \citenamefont {Zhang}, \citenamefont {Park},
  \citenamefont {Yu}, \citenamefont {Li}, \citenamefont {Ma}, \citenamefont
  {Wang}, \citenamefont {Huang}, \citenamefont {Song} \emph
  {et~al.}}]{zhang2022recent}%
  \BibitemOpen
  \bibfield  {author} {\bibinfo {author} {\bibfnamefont {J.}~\bibnamefont
  {Zhang}}, \bibinfo {author} {\bibfnamefont {Z.}~\bibnamefont {Zhou}},
  \bibinfo {author} {\bibfnamefont {Z.}~\bibnamefont {Zhang}}, \bibinfo
  {author} {\bibfnamefont {M.}~\bibnamefont {Park}}, \bibinfo {author}
  {\bibfnamefont {Q.}~\bibnamefont {Yu}}, \bibinfo {author} {\bibfnamefont
  {Z.}~\bibnamefont {Li}}, \bibinfo {author} {\bibfnamefont {J.}~\bibnamefont
  {Ma}}, \bibinfo {author} {\bibfnamefont {A.}~\bibnamefont {Wang}}, \bibinfo
  {author} {\bibfnamefont {H.}~\bibnamefont {Huang}}, \bibinfo {author}
  {\bibfnamefont {M.}~\bibnamefont {Song}}, \emph {et~al.},\ }\bibfield
  {title} {\bibinfo {title} {Recent development of chemically complex metallic
  glasses: from accelerated compositional design, additive manufacturing to
  novel applications},\ }\href@noop {} {\bibfield  {journal} {\bibinfo
  {journal} {Materials Futures}\ }\textbf {\bibinfo {volume} {1}},\ \bibinfo
  {pages} {012001} (\bibinfo {year} {2022})}\BibitemShut {NoStop}%
\bibitem [{\citenamefont {Zhou}\ \emph {et~al.}(2023)\citenamefont {Zhou},
  \citenamefont {Shang}, \citenamefont {Liu},\ and\ \citenamefont
  {Yang}}]{zhou2023generative}%
  \BibitemOpen
  \bibfield  {author} {\bibinfo {author} {\bibfnamefont {Z.}~\bibnamefont
  {Zhou}}, \bibinfo {author} {\bibfnamefont {Y.}~\bibnamefont {Shang}},
  \bibinfo {author} {\bibfnamefont {X.}~\bibnamefont {Liu}},\ and\ \bibinfo
  {author} {\bibfnamefont {Y.}~\bibnamefont {Yang}},\ }\bibfield  {title}
  {\bibinfo {title} {A generative deep learning framework for inverse design of
  compositionally complex bulk metallic glasses},\ }\href@noop {} {\bibfield
  {journal} {\bibinfo  {journal} {npj Computational Materials}\ }\textbf
  {\bibinfo {volume} {9}},\ \bibinfo {pages} {15} (\bibinfo {year}
  {2023})}\BibitemShut {NoStop}%
\bibitem [{\citenamefont {Cai}\ \emph {et~al.}(2024)\citenamefont {Cai},
  \citenamefont {Cai}, \citenamefont {Yang}, \citenamefont {Wang},
  \citenamefont {Yao}, \citenamefont {Gao}, \citenamefont {Chang},
  \citenamefont {Li}, \citenamefont {Xu}, \citenamefont {Wang} \emph
  {et~al.}}]{cai2024uni}%
  \BibitemOpen
  \bibfield  {author} {\bibinfo {author} {\bibfnamefont {H.}~\bibnamefont
  {Cai}}, \bibinfo {author} {\bibfnamefont {X.}~\bibnamefont {Cai}}, \bibinfo
  {author} {\bibfnamefont {S.}~\bibnamefont {Yang}}, \bibinfo {author}
  {\bibfnamefont {J.}~\bibnamefont {Wang}}, \bibinfo {author} {\bibfnamefont
  {L.}~\bibnamefont {Yao}}, \bibinfo {author} {\bibfnamefont {Z.}~\bibnamefont
  {Gao}}, \bibinfo {author} {\bibfnamefont {J.}~\bibnamefont {Chang}}, \bibinfo
  {author} {\bibfnamefont {S.}~\bibnamefont {Li}}, \bibinfo {author}
  {\bibfnamefont {M.}~\bibnamefont {Xu}}, \bibinfo {author} {\bibfnamefont
  {C.}~\bibnamefont {Wang}}, \emph {et~al.},\ }\bibfield  {title} {\bibinfo
  {title} {Uni-smart: Universal science multimodal analysis and research
  transformer},\ }\href@noop {} {\bibfield  {journal} {\bibinfo  {journal}
  {arXiv preprint arXiv:2403.10301}\ } (\bibinfo {year} {2024})}\BibitemShut
  {NoStop}%
\bibitem [{\citenamefont {Zheng}\ \emph {et~al.}(2023)\citenamefont {Zheng},
  \citenamefont {Zhang}, \citenamefont {Borgs}, \citenamefont {Chayes},\ and\
  \citenamefont {Yaghi}}]{zheng2023chatgpt}%
  \BibitemOpen
  \bibfield  {author} {\bibinfo {author} {\bibfnamefont {Z.}~\bibnamefont
  {Zheng}}, \bibinfo {author} {\bibfnamefont {O.}~\bibnamefont {Zhang}},
  \bibinfo {author} {\bibfnamefont {C.}~\bibnamefont {Borgs}}, \bibinfo
  {author} {\bibfnamefont {J.~T.}\ \bibnamefont {Chayes}},\ and\ \bibinfo
  {author} {\bibfnamefont {O.~M.}\ \bibnamefont {Yaghi}},\ }\bibfield  {title}
  {\bibinfo {title} {Chatgpt chemistry assistant for text mining and the
  prediction of mof synthesis},\ }\href@noop {} {\bibfield  {journal} {\bibinfo
   {journal} {Journal of the American Chemical Society}\ }\textbf {\bibinfo
  {volume} {145}},\ \bibinfo {pages} {18048} (\bibinfo {year}
  {2023})}\BibitemShut {NoStop}%
\bibitem [{\citenamefont {Mirza}\ and\ \citenamefont
  {Osindero}(2014)}]{mirza2014conditional}%
  \BibitemOpen
  \bibfield  {author} {\bibinfo {author} {\bibfnamefont {M.}~\bibnamefont
  {Mirza}}\ and\ \bibinfo {author} {\bibfnamefont {S.}~\bibnamefont
  {Osindero}},\ }\bibfield  {title} {\bibinfo {title} {Conditional generative
  adversarial nets},\ }\href@noop {} {\bibfield  {journal} {\bibinfo  {journal}
  {arXiv preprint arXiv:1411.1784}\ } (\bibinfo {year} {2014})}\BibitemShut
  {NoStop}%
\bibitem [{\citenamefont {Yang}\ \emph {et~al.}(2004)\citenamefont {Yang},
  \citenamefont {Zhang}, \citenamefont {Frangi},\ and\ \citenamefont
  {Yang}}]{yang2004two}%
  \BibitemOpen
  \bibfield  {author} {\bibinfo {author} {\bibfnamefont {J.}~\bibnamefont
  {Yang}}, \bibinfo {author} {\bibfnamefont {D.}~\bibnamefont {Zhang}},
  \bibinfo {author} {\bibfnamefont {A.~F.}\ \bibnamefont {Frangi}},\ and\
  \bibinfo {author} {\bibfnamefont {J.-y.}\ \bibnamefont {Yang}},\ }\bibfield
  {title} {\bibinfo {title} {Two-dimensional pca: a new approach to
  appearance-based face representation and recognition},\ }\href@noop {}
  {\bibfield  {journal} {\bibinfo  {journal} {IEEE transactions on pattern
  analysis and machine intelligence}\ }\textbf {\bibinfo {volume} {26}},\
  \bibinfo {pages} {131} (\bibinfo {year} {2004})}\BibitemShut {NoStop}%
\bibitem [{\citenamefont {Kullback}\ and\ \citenamefont
  {Leibler}(1951)}]{kullback1951information}%
  \BibitemOpen
  \bibfield  {author} {\bibinfo {author} {\bibfnamefont {S.}~\bibnamefont
  {Kullback}}\ and\ \bibinfo {author} {\bibfnamefont {R.~A.}\ \bibnamefont
  {Leibler}},\ }\bibfield  {title} {\bibinfo {title} {On information and
  sufficiency},\ }\href@noop {} {\bibfield  {journal} {\bibinfo  {journal} {The
  annals of mathematical statistics}\ }\textbf {\bibinfo {volume} {22}},\
  \bibinfo {pages} {79} (\bibinfo {year} {1951})}\BibitemShut {NoStop}%
\bibitem [{\citenamefont {Tishby}\ \emph {et~al.}(2000)\citenamefont {Tishby},
  \citenamefont {Pereira},\ and\ \citenamefont
  {Bialek}}]{tishby2000information}%
  \BibitemOpen
  \bibfield  {author} {\bibinfo {author} {\bibfnamefont {N.}~\bibnamefont
  {Tishby}}, \bibinfo {author} {\bibfnamefont {F.~C.}\ \bibnamefont
  {Pereira}},\ and\ \bibinfo {author} {\bibfnamefont {W.}~\bibnamefont
  {Bialek}},\ }\bibfield  {title} {\bibinfo {title} {The information bottleneck
  method},\ }\href@noop {} {\bibfield  {journal} {\bibinfo  {journal} {arXiv
  preprint physics/0004057}\ } (\bibinfo {year} {2000})}\BibitemShut {NoStop}%
\bibitem [{\citenamefont {Vaserstein}(1969)}]{vaserstein1969markov}%
  \BibitemOpen
  \bibfield  {author} {\bibinfo {author} {\bibfnamefont {L.~N.}\ \bibnamefont
  {Vaserstein}},\ }\bibfield  {title} {\bibinfo {title} {Markov processes over
  denumerable products of spaces, describing large systems of automata},\
  }\href@noop {} {\bibfield  {journal} {\bibinfo  {journal} {Problemy Peredachi
  Informatsii}\ }\textbf {\bibinfo {volume} {5}},\ \bibinfo {pages} {64}
  (\bibinfo {year} {1969})}\BibitemShut {NoStop}%
\bibitem [{\citenamefont {LeCun}\ \emph {et~al.}(2015)\citenamefont {LeCun},
  \citenamefont {Bengio},\ and\ \citenamefont {Hinton}}]{lecun2015deep}%
  \BibitemOpen
  \bibfield  {author} {\bibinfo {author} {\bibfnamefont {Y.}~\bibnamefont
  {LeCun}}, \bibinfo {author} {\bibfnamefont {Y.}~\bibnamefont {Bengio}},\ and\
  \bibinfo {author} {\bibfnamefont {G.}~\bibnamefont {Hinton}},\ }\bibfield
  {title} {\bibinfo {title} {Deep learning},\ }\href@noop {} {\bibfield
  {journal} {\bibinfo  {journal} {Nature}\ }\textbf {\bibinfo {volume} {521}},\
  \bibinfo {pages} {436} (\bibinfo {year} {2015})}\BibitemShut {NoStop}%
\bibitem [{\citenamefont {Lu}\ \emph {et~al.}(2000)\citenamefont {Lu},
  \citenamefont {Tan}, \citenamefont {Ng},\ and\ \citenamefont
  {Li}}]{lu2000correlation}%
  \BibitemOpen
  \bibfield  {author} {\bibinfo {author} {\bibfnamefont {Z.}~\bibnamefont
  {Lu}}, \bibinfo {author} {\bibfnamefont {H.}~\bibnamefont {Tan}}, \bibinfo
  {author} {\bibfnamefont {S.}~\bibnamefont {Ng}},\ and\ \bibinfo {author}
  {\bibfnamefont {Y.}~\bibnamefont {Li}},\ }\bibfield  {title} {\bibinfo
  {title} {The correlation between reduced glass transition temperature and
  glass forming ability of bulk metallic glasses},\ }\href@noop {} {\bibfield
  {journal} {\bibinfo  {journal} {Scripta materialia}\ }\textbf {\bibinfo
  {volume} {42}},\ \bibinfo {pages} {667} (\bibinfo {year} {2000})}\BibitemShut
  {NoStop}%
\bibitem [{\citenamefont {Lu}\ and\ \citenamefont {Liu}(2002)}]{lu2002new}%
  \BibitemOpen
  \bibfield  {author} {\bibinfo {author} {\bibfnamefont {Z.}~\bibnamefont
  {Lu}}\ and\ \bibinfo {author} {\bibfnamefont {C.}~\bibnamefont {Liu}},\
  }\bibfield  {title} {\bibinfo {title} {A new glass-forming ability criterion
  for bulk metallic glasses},\ }\href@noop {} {\bibfield  {journal} {\bibinfo
  {journal} {Acta materialia}\ }\textbf {\bibinfo {volume} {50}},\ \bibinfo
  {pages} {3501} (\bibinfo {year} {2002})}\BibitemShut {NoStop}%
\bibitem [{\citenamefont {{Dennis Bastian}}(2019)}]{cost_report}%
  \BibitemOpen
  \bibfield  {author} {\bibinfo {author} {\bibnamefont {{Dennis Bastian}}},\
  }\href@noop {} {\bibinfo {title} {Preismonitor dezember 2019}} (\bibinfo
  {year} {2019}),\ \bibinfo {note} {bundesanstalt fürGeowissenschaftenund
  Rohstoffe,
  \url{https://www.bgr.bund.de/DE/Themen/Min_rohstoffe/Produkte/Preisliste/pm_19_12.pdf}}\BibitemShut
  {NoStop}%
\bibitem [{\citenamefont {Cullity}\ and\ \citenamefont
  {Stock}(2001)}]{book_xrd}%
  \BibitemOpen
  \bibfield  {author} {\bibinfo {author} {\bibfnamefont {B.~D.}\ \bibnamefont
  {Cullity}}\ and\ \bibinfo {author} {\bibfnamefont {S.~R.}\ \bibnamefont
  {Stock}},\ }\href@noop {} {\emph {\bibinfo {title} {Elements of X-ray
  Diffraction}}},\ \bibinfo {edition} {3rd}\ ed.\ (\bibinfo  {publisher}
  {Prentice Hall},\ \bibinfo {address} {Upper Saddle River, NJ},\ \bibinfo
  {year} {2001})\BibitemShut {NoStop}%
\bibitem [{\citenamefont {Richardson}\ and\ \citenamefont
  {Tanner}(1997)}]{book_dsc}%
  \BibitemOpen
  \bibfield  {author} {\bibinfo {author} {\bibfnamefont {M.~J.}\ \bibnamefont
  {Richardson}}\ and\ \bibinfo {author} {\bibfnamefont {K.~E.}\ \bibnamefont
  {Tanner}},\ }\href@noop {} {\emph {\bibinfo {title} {Principles of Thermal
  Analysis and Calorimetry}}}\ (\bibinfo  {publisher} {Royal Society of
  Chemistry},\ \bibinfo {address} {Cambridge},\ \bibinfo {year} {1997})\
  Chap.~\bibinfo {chapter} {4}, pp.\ \bibinfo {pages} {95--123}\BibitemShut
  {NoStop}%
\bibitem [{\citenamefont {Greer}\ and\ \citenamefont
  {Lu}(2015)}]{greer2015metallic}%
  \BibitemOpen
  \bibfield  {author} {\bibinfo {author} {\bibfnamefont {A.~L.}\ \bibnamefont
  {Greer}}\ and\ \bibinfo {author} {\bibfnamefont {Z.}~\bibnamefont {Lu}},\
  }\bibfield  {title} {\bibinfo {title} {Thermal analysis of metallic glasses:
  Glass transition and crystallization},\ }\href@noop {} {\bibfield  {journal}
  {\bibinfo  {journal} {Progress in Materials Science}\ }\textbf {\bibinfo
  {volume} {74}},\ \bibinfo {pages} {71} (\bibinfo {year} {2015})}\BibitemShut
  {NoStop}%
\bibitem [{\citenamefont {Liu}\ \emph {et~al.}(2024)\citenamefont {Liu},
  \citenamefont {Wen}, \citenamefont {Pattamatta},\ and\ \citenamefont
  {Srolovitz}}]{liu2024}%
  \BibitemOpen
  \bibfield  {author} {\bibinfo {author} {\bibfnamefont {S.}~\bibnamefont
  {Liu}}, \bibinfo {author} {\bibfnamefont {T.}~\bibnamefont {Wen}}, \bibinfo
  {author} {\bibfnamefont {A.~S.}\ \bibnamefont {Pattamatta}},\ and\ \bibinfo
  {author} {\bibfnamefont {D.~J.}\ \bibnamefont {Srolovitz}},\ }\bibfield
  {title} {\bibinfo {title} {A prompt-engineered large language model, deep
  learning workflow for materials classification},\ }\href@noop {} {\bibfield
  {journal} {\bibinfo  {journal} {Materials Today}\ }\textbf {\bibinfo {volume}
  {80}},\ \bibinfo {pages} {240} (\bibinfo {year} {2024})}\BibitemShut
  {NoStop}%
\bibitem [{\citenamefont {Hu}\ \emph {et~al.}(2024)\citenamefont {Hu},
  \citenamefont {Liu}, \citenamefont {Ye}, \citenamefont {Hao},\ and\
  \citenamefont {Wen}}]{hu2024}%
  \BibitemOpen
  \bibfield  {author} {\bibinfo {author} {\bibfnamefont {B.}~\bibnamefont
  {Hu}}, \bibinfo {author} {\bibfnamefont {S.}~\bibnamefont {Liu}}, \bibinfo
  {author} {\bibfnamefont {B.}~\bibnamefont {Ye}}, \bibinfo {author}
  {\bibfnamefont {Y.}~\bibnamefont {Hao}},\ and\ \bibinfo {author}
  {\bibfnamefont {T.}~\bibnamefont {Wen}},\ }\href@noop {} {\bibinfo {title} {A
  multi-agent framework for materials laws discovery}} (\bibinfo {year}
  {2024})\BibitemShut {NoStop}%
\bibitem [{\citenamefont {Goodfellow}\ \emph {et~al.}(2014)\citenamefont
  {Goodfellow}, \citenamefont {Pouget-Abadie}, \citenamefont {Mirza},
  \citenamefont {Xu}, \citenamefont {Warde-Farley}, \citenamefont {Ozair},
  \citenamefont {Courville},\ and\ \citenamefont
  {Bengio}}]{goodfellow2014generative}%
  \BibitemOpen
  \bibfield  {author} {\bibinfo {author} {\bibfnamefont {I.~J.}\ \bibnamefont
  {Goodfellow}}, \bibinfo {author} {\bibfnamefont {J.}~\bibnamefont
  {Pouget-Abadie}}, \bibinfo {author} {\bibfnamefont {M.}~\bibnamefont
  {Mirza}}, \bibinfo {author} {\bibfnamefont {B.}~\bibnamefont {Xu}}, \bibinfo
  {author} {\bibfnamefont {D.}~\bibnamefont {Warde-Farley}}, \bibinfo {author}
  {\bibfnamefont {S.}~\bibnamefont {Ozair}}, \bibinfo {author} {\bibfnamefont
  {A.}~\bibnamefont {Courville}},\ and\ \bibinfo {author} {\bibfnamefont
  {Y.}~\bibnamefont {Bengio}},\ }\href@noop {} {\bibinfo {title} {Generative
  adversarial networks}} (\bibinfo {year} {2014})\BibitemShut {NoStop}%
\bibitem [{\citenamefont {Nash}\ and\ \citenamefont
  {Sutcliffe}(1970)}]{nash1970}%
  \BibitemOpen
  \bibfield  {author} {\bibinfo {author} {\bibfnamefont {J.}~\bibnamefont
  {Nash}}\ and\ \bibinfo {author} {\bibfnamefont {J.}~\bibnamefont
  {Sutcliffe}},\ }\bibfield  {title} {\bibinfo {title} {River flow forecasting
  through conceptual models part i — a discussion of principles},\
  }\href@noop {} {\bibfield  {journal} {\bibinfo  {journal} {Journal of
  Hydrology}\ }\textbf {\bibinfo {volume} {10}},\ \bibinfo {pages} {282}
  (\bibinfo {year} {1970})}\BibitemShut {NoStop}%
\bibitem [{\citenamefont {Inoue}\ \emph {et~al.}(1991)\citenamefont {Inoue},
  \citenamefont {Kato}, \citenamefont {Zhang},\ and\ \citenamefont
  {Masumoto}}]{inoue1991mg}%
  \BibitemOpen
  \bibfield  {author} {\bibinfo {author} {\bibfnamefont {A.}~\bibnamefont
  {Inoue}}, \bibinfo {author} {\bibfnamefont {A.}~\bibnamefont {Kato}},
  \bibinfo {author} {\bibfnamefont {T.}~\bibnamefont {Zhang}},\ and\ \bibinfo
  {author} {\bibfnamefont {T.}~\bibnamefont {Masumoto}},\ }\bibfield  {title}
  {\bibinfo {title} {Mg--cu--y amorphous alloys with high mechanical strengths
  produced by a metallic mold casting method},\ }\href@noop {} {\bibfield
  {journal} {\bibinfo  {journal} {Materials transactions, JIM}\ }\textbf
  {\bibinfo {volume} {32}},\ \bibinfo {pages} {609} (\bibinfo {year}
  {1991})}\BibitemShut {NoStop}%
\bibitem [{\citenamefont {Xiao}\ \emph {et~al.}(2004)\citenamefont {Xiao},
  \citenamefont {Shoushi}, \citenamefont {Guoming}, \citenamefont {Qin},\ and\
  \citenamefont {Yuanda}}]{xiao2004influence}%
  \BibitemOpen
  \bibfield  {author} {\bibinfo {author} {\bibfnamefont {X.}~\bibnamefont
  {Xiao}}, \bibinfo {author} {\bibfnamefont {F.}~\bibnamefont {Shoushi}},
  \bibinfo {author} {\bibfnamefont {W.}~\bibnamefont {Guoming}}, \bibinfo
  {author} {\bibfnamefont {H.}~\bibnamefont {Qin}},\ and\ \bibinfo {author}
  {\bibfnamefont {D.}~\bibnamefont {Yuanda}},\ }\bibfield  {title} {\bibinfo
  {title} {Influence of beryllium on thermal stability and glass-forming
  ability of zr--al--ni--cu bulk amorphous alloys},\ }\href@noop {} {\bibfield
  {journal} {\bibinfo  {journal} {Journal of alloys and compounds}\ }\textbf
  {\bibinfo {volume} {376}},\ \bibinfo {pages} {145} (\bibinfo {year}
  {2004})}\BibitemShut {NoStop}%
\bibitem [{\citenamefont {Mondal}\ and\ \citenamefont
  {Murty}(2005)}]{mondal2005parameters}%
  \BibitemOpen
  \bibfield  {author} {\bibinfo {author} {\bibfnamefont {K.}~\bibnamefont
  {Mondal}}\ and\ \bibinfo {author} {\bibfnamefont {B.}~\bibnamefont {Murty}},\
  }\bibfield  {title} {\bibinfo {title} {On the parameters to assess the glass
  forming ability of liquids},\ }\href@noop {} {\bibfield  {journal} {\bibinfo
  {journal} {Journal of non-crystalline solids}\ }\textbf {\bibinfo {volume}
  {351}},\ \bibinfo {pages} {1366} (\bibinfo {year} {2005})}\BibitemShut
  {NoStop}%
\bibitem [{\citenamefont {Chen}\ \emph {et~al.}(2006)\citenamefont {Chen},
  \citenamefont {Shen}, \citenamefont {Zhang}, \citenamefont {Fan},
  \citenamefont {Sun},\ and\ \citenamefont {McCartney}}]{chen2006new}%
  \BibitemOpen
  \bibfield  {author} {\bibinfo {author} {\bibfnamefont {Q.}~\bibnamefont
  {Chen}}, \bibinfo {author} {\bibfnamefont {J.}~\bibnamefont {Shen}}, \bibinfo
  {author} {\bibfnamefont {D.}~\bibnamefont {Zhang}}, \bibinfo {author}
  {\bibfnamefont {H.}~\bibnamefont {Fan}}, \bibinfo {author} {\bibfnamefont
  {J.}~\bibnamefont {Sun}},\ and\ \bibinfo {author} {\bibfnamefont
  {D.}~\bibnamefont {McCartney}},\ }\bibfield  {title} {\bibinfo {title} {A new
  criterion for evaluating the glass-forming ability of bulk metallic
  glasses},\ }\href@noop {} {\bibfield  {journal} {\bibinfo  {journal}
  {Materials Science and Engineering: A}\ }\textbf {\bibinfo {volume} {433}},\
  \bibinfo {pages} {155} (\bibinfo {year} {2006})}\BibitemShut {NoStop}%
\bibitem [{\citenamefont {Du}\ \emph {et~al.}(2007)\citenamefont {Du},
  \citenamefont {Huang}, \citenamefont {Liu},\ and\ \citenamefont
  {Lu}}]{du2007new}%
  \BibitemOpen
  \bibfield  {author} {\bibinfo {author} {\bibfnamefont {X.}~\bibnamefont
  {Du}}, \bibinfo {author} {\bibfnamefont {J.}~\bibnamefont {Huang}}, \bibinfo
  {author} {\bibfnamefont {C.}~\bibnamefont {Liu}},\ and\ \bibinfo {author}
  {\bibfnamefont {Z.}~\bibnamefont {Lu}},\ }\bibfield  {title} {\bibinfo
  {title} {New criterion of glass forming ability for bulk metallic glasses},\
  }\href@noop {} {\bibfield  {journal} {\bibinfo  {journal} {Journal of applied
  physics}\ }\textbf {\bibinfo {volume} {101}} (\bibinfo {year}
  {2007})}\BibitemShut {NoStop}%
\bibitem [{\citenamefont {Fan}\ \emph {et~al.}(2007)\citenamefont {Fan},
  \citenamefont {Choo},\ and\ \citenamefont {Liaw}}]{fan2007new}%
  \BibitemOpen
  \bibfield  {author} {\bibinfo {author} {\bibfnamefont {G.}~\bibnamefont
  {Fan}}, \bibinfo {author} {\bibfnamefont {H.}~\bibnamefont {Choo}},\ and\
  \bibinfo {author} {\bibfnamefont {P.}~\bibnamefont {Liaw}},\ }\bibfield
  {title} {\bibinfo {title} {A new criterion for the glass-forming ability of
  liquids},\ }\href@noop {} {\bibfield  {journal} {\bibinfo  {journal} {Journal
  of Non-Crystalline Solids}\ }\textbf {\bibinfo {volume} {353}},\ \bibinfo
  {pages} {102} (\bibinfo {year} {2007})}\BibitemShut {NoStop}%
\bibitem [{\citenamefont {Du}\ and\ \citenamefont {Huang}(2008)}]{du2008new}%
  \BibitemOpen
  \bibfield  {author} {\bibinfo {author} {\bibfnamefont {X.}~\bibnamefont
  {Du}}\ and\ \bibinfo {author} {\bibfnamefont {J.}~\bibnamefont {Huang}},\
  }\bibfield  {title} {\bibinfo {title} {New criterion in predicting glass
  forming ability of various glass-forming systems},\ }\href@noop {} {\bibfield
   {journal} {\bibinfo  {journal} {Chinese Physics B}\ }\textbf {\bibinfo
  {volume} {17}},\ \bibinfo {pages} {249} (\bibinfo {year} {2008})}\BibitemShut
  {NoStop}%
\bibitem [{\citenamefont {Yuan}\ \emph {et~al.}(2008)\citenamefont {Yuan},
  \citenamefont {Bao}, \citenamefont {Lu}, \citenamefont {Zhang},\ and\
  \citenamefont {Yao}}]{yuan2008new}%
  \BibitemOpen
  \bibfield  {author} {\bibinfo {author} {\bibfnamefont {Z.-Z.}\ \bibnamefont
  {Yuan}}, \bibinfo {author} {\bibfnamefont {S.-L.}\ \bibnamefont {Bao}},
  \bibinfo {author} {\bibfnamefont {Y.}~\bibnamefont {Lu}}, \bibinfo {author}
  {\bibfnamefont {D.-P.}\ \bibnamefont {Zhang}},\ and\ \bibinfo {author}
  {\bibfnamefont {L.}~\bibnamefont {Yao}},\ }\bibfield  {title} {\bibinfo
  {title} {A new criterion for evaluating the glass-forming ability of bulk
  glass forming alloys},\ }\href@noop {} {\bibfield  {journal} {\bibinfo
  {journal} {Journal of Alloys and Compounds}\ }\textbf {\bibinfo {volume}
  {459}},\ \bibinfo {pages} {251} (\bibinfo {year} {2008})}\BibitemShut
  {NoStop}%
\bibitem [{\citenamefont {Long}\ \emph {et~al.}(2009)\citenamefont {Long},
  \citenamefont {Wei}, \citenamefont {Ding}, \citenamefont {Zhang},
  \citenamefont {Xie},\ and\ \citenamefont {Inoue}}]{long2009new}%
  \BibitemOpen
  \bibfield  {author} {\bibinfo {author} {\bibfnamefont {Z.}~\bibnamefont
  {Long}}, \bibinfo {author} {\bibfnamefont {H.}~\bibnamefont {Wei}}, \bibinfo
  {author} {\bibfnamefont {Y.}~\bibnamefont {Ding}}, \bibinfo {author}
  {\bibfnamefont {P.}~\bibnamefont {Zhang}}, \bibinfo {author} {\bibfnamefont
  {G.}~\bibnamefont {Xie}},\ and\ \bibinfo {author} {\bibfnamefont
  {A.}~\bibnamefont {Inoue}},\ }\bibfield  {title} {\bibinfo {title} {A new
  criterion for predicting the glass-forming ability of bulk metallic
  glasses},\ }\href@noop {} {\bibfield  {journal} {\bibinfo  {journal} {Journal
  of Alloys and Compounds}\ }\textbf {\bibinfo {volume} {475}},\ \bibinfo
  {pages} {207} (\bibinfo {year} {2009})}\BibitemShut {NoStop}%
\bibitem [{\citenamefont {JI}\ and\ \citenamefont
  {Ye}(2009)}]{ji2009thermodynamic}%
  \BibitemOpen
  \bibfield  {author} {\bibinfo {author} {\bibfnamefont {X.-l.}\ \bibnamefont
  {JI}}\ and\ \bibinfo {author} {\bibfnamefont {P.}~\bibnamefont {Ye}},\
  }\bibfield  {title} {\bibinfo {title} {A thermodynamic approach to assess
  glass-forming ability of bulk metallic glasses},\ }\href@noop {} {\bibfield
  {journal} {\bibinfo  {journal} {Transactions of Nonferrous Metals Society of
  China}\ }\textbf {\bibinfo {volume} {19}},\ \bibinfo {pages} {1271} (\bibinfo
  {year} {2009})}\BibitemShut {NoStop}%
\bibitem [{\citenamefont {Zhang}\ and\ \citenamefont
  {Chou}(2009)}]{zhang2009criterion}%
  \BibitemOpen
  \bibfield  {author} {\bibinfo {author} {\bibfnamefont {G.-H.}\ \bibnamefont
  {Zhang}}\ and\ \bibinfo {author} {\bibfnamefont {K.-C.}\ \bibnamefont
  {Chou}},\ }\bibfield  {title} {\bibinfo {title} {A criterion for evaluating
  glass-forming ability of alloys},\ }\href@noop {} {\bibfield  {journal}
  {\bibinfo  {journal} {Journal of Applied Physics}\ }\textbf {\bibinfo
  {volume} {106}} (\bibinfo {year} {2009})}\BibitemShut {NoStop}%
\bibitem [{\citenamefont {Hongqing}\ \emph {et~al.}(2009)\citenamefont
  {Hongqing}, \citenamefont {Xiangan}, \citenamefont {Zhilin}, \citenamefont
  {Jian}, \citenamefont {Ping},\ and\ \citenamefont
  {Zhichun}}]{2009Correlations}%
  \BibitemOpen
  \bibfield  {author} {\bibinfo {author} {\bibfnamefont {W.}~\bibnamefont
  {Hongqing}}, \bibinfo {author} {\bibfnamefont {L.}~\bibnamefont {Xiangan}},
  \bibinfo {author} {\bibfnamefont {L.}~\bibnamefont {Zhilin}}, \bibinfo
  {author} {\bibfnamefont {P.}~\bibnamefont {Jian}}, \bibinfo {author}
  {\bibfnamefont {Z.}~\bibnamefont {Ping}},\ and\ \bibinfo {author}
  {\bibfnamefont {Z.}~\bibnamefont {Zhichun}},\ }\bibfield  {title} {\bibinfo
  {title} {Correlations between viscosity and glass-forming ability in bulk
  amorphous alloys},\ }\href@noop {} {\bibfield  {journal} {\bibinfo  {journal}
  {Acta Physica Sinica}\ }\textbf {\bibinfo {volume} {58}} (\bibinfo {year}
  {2009})}\BibitemShut {NoStop}%
\bibitem [{\citenamefont {Guo}\ and\ \citenamefont {Liu}(2010)}]{guo2010new}%
  \BibitemOpen
  \bibfield  {author} {\bibinfo {author} {\bibfnamefont {S.}~\bibnamefont
  {Guo}}\ and\ \bibinfo {author} {\bibfnamefont {C.}~\bibnamefont {Liu}},\
  }\bibfield  {title} {\bibinfo {title} {New glass forming ability criterion
  derived from cooling consideration},\ }\href@noop {} {\bibfield  {journal}
  {\bibinfo  {journal} {Intermetallics}\ }\textbf {\bibinfo {volume} {18}},\
  \bibinfo {pages} {2065} (\bibinfo {year} {2010})}\BibitemShut {NoStop}%
\bibitem [{\citenamefont {Dong}\ \emph {et~al.}(2011)\citenamefont {Dong},
  \citenamefont {Zhou}, \citenamefont {Li}, \citenamefont {Lu}, \citenamefont
  {Feng}, \citenamefont {Ni},\ and\ \citenamefont {Lu}}]{dong2011new}%
  \BibitemOpen
  \bibfield  {author} {\bibinfo {author} {\bibfnamefont {B.-s.}\ \bibnamefont
  {Dong}}, \bibinfo {author} {\bibfnamefont {S.-x.}\ \bibnamefont {Zhou}},
  \bibinfo {author} {\bibfnamefont {D.-r.}\ \bibnamefont {Li}}, \bibinfo
  {author} {\bibfnamefont {C.-w.}\ \bibnamefont {Lu}}, \bibinfo {author}
  {\bibfnamefont {G.}~\bibnamefont {Feng}}, \bibinfo {author} {\bibfnamefont
  {X.-j.}\ \bibnamefont {Ni}},\ and\ \bibinfo {author} {\bibfnamefont {Z.-c.}\
  \bibnamefont {Lu}},\ }\bibfield  {title} {\bibinfo {title} {A new criterion
  for predicting glass forming ability of bulk metallic glasses and some
  critical discussions},\ }\href@noop {} {\bibfield  {journal} {\bibinfo
  {journal} {Progress in natural science: Materials international}\ }\textbf
  {\bibinfo {volume} {21}},\ \bibinfo {pages} {164} (\bibinfo {year}
  {2011})}\BibitemShut {NoStop}%
\bibitem [{\citenamefont {B{\l}yskun}\ \emph {et~al.}(2015)\citenamefont
  {B{\l}yskun}, \citenamefont {Maj}, \citenamefont {Kowalczyk}, \citenamefont
  {Latuch},\ and\ \citenamefont {Kulik}}]{blyskun2015relation}%
  \BibitemOpen
  \bibfield  {author} {\bibinfo {author} {\bibfnamefont {P.}~\bibnamefont
  {B{\l}yskun}}, \bibinfo {author} {\bibfnamefont {P.}~\bibnamefont {Maj}},
  \bibinfo {author} {\bibfnamefont {M.}~\bibnamefont {Kowalczyk}}, \bibinfo
  {author} {\bibfnamefont {J.}~\bibnamefont {Latuch}},\ and\ \bibinfo {author}
  {\bibfnamefont {T.}~\bibnamefont {Kulik}},\ }\bibfield  {title} {\bibinfo
  {title} {Relation of various gfa indicators to the critical diameter of
  zr-based bmgs},\ }\href@noop {} {\bibfield  {journal} {\bibinfo  {journal}
  {Journal of Alloys and Compounds}\ }\textbf {\bibinfo {volume} {625}},\
  \bibinfo {pages} {13} (\bibinfo {year} {2015})}\BibitemShut {NoStop}%
\bibitem [{\citenamefont {Tripathi}\ \emph {et~al.}(2016)\citenamefont
  {Tripathi}, \citenamefont {Ganguly}, \citenamefont {Dey},\ and\ \citenamefont
  {Chattopadhyay}}]{tripathi2016evolution}%
  \BibitemOpen
  \bibfield  {author} {\bibinfo {author} {\bibfnamefont {M.~K.}\ \bibnamefont
  {Tripathi}}, \bibinfo {author} {\bibfnamefont {S.}~\bibnamefont {Ganguly}},
  \bibinfo {author} {\bibfnamefont {P.}~\bibnamefont {Dey}},\ and\ \bibinfo
  {author} {\bibfnamefont {P.}~\bibnamefont {Chattopadhyay}},\ }\bibfield
  {title} {\bibinfo {title} {Evolution of glass forming ability indicator by
  genetic programming},\ }\href@noop {} {\bibfield  {journal} {\bibinfo
  {journal} {Computational Materials Science}\ }\textbf {\bibinfo {volume}
  {118}},\ \bibinfo {pages} {56} (\bibinfo {year} {2016})}\BibitemShut
  {NoStop}%
\bibitem [{\citenamefont {Long}\ \emph {et~al.}(2018)\citenamefont {Long},
  \citenamefont {Liu}, \citenamefont {Zhong}, \citenamefont {Zhang},
  \citenamefont {Zhao}, \citenamefont {Liao},\ and\ \citenamefont
  {Chen}}]{long2018new}%
  \BibitemOpen
  \bibfield  {author} {\bibinfo {author} {\bibfnamefont {Z.}~\bibnamefont
  {Long}}, \bibinfo {author} {\bibfnamefont {W.}~\bibnamefont {Liu}}, \bibinfo
  {author} {\bibfnamefont {M.}~\bibnamefont {Zhong}}, \bibinfo {author}
  {\bibfnamefont {Y.}~\bibnamefont {Zhang}}, \bibinfo {author} {\bibfnamefont
  {M.}~\bibnamefont {Zhao}}, \bibinfo {author} {\bibfnamefont {G.}~\bibnamefont
  {Liao}},\ and\ \bibinfo {author} {\bibfnamefont {Z.}~\bibnamefont {Chen}},\
  }\bibfield  {title} {\bibinfo {title} {A new correlation between the
  characteristics temperature and glass-forming ability for bulk metallic
  glasses},\ }\href@noop {} {\bibfield  {journal} {\bibinfo  {journal} {Journal
  of Thermal Analysis and Calorimetry}\ }\textbf {\bibinfo {volume} {132}},\
  \bibinfo {pages} {1645} (\bibinfo {year} {2018})}\BibitemShut {NoStop}%
\bibitem [{\citenamefont {Deng}\ and\ \citenamefont
  {Zhang}(2020)}]{deng2020critical}%
  \BibitemOpen
  \bibfield  {author} {\bibinfo {author} {\bibfnamefont {B.}~\bibnamefont
  {Deng}}\ and\ \bibinfo {author} {\bibfnamefont {Y.}~\bibnamefont {Zhang}},\
  }\bibfield  {title} {\bibinfo {title} {Critical feature space for predicting
  the glass forming ability of metallic alloys revealed by machine learning},\
  }\href@noop {} {\bibfield  {journal} {\bibinfo  {journal} {Chemical Physics}\
  }\textbf {\bibinfo {volume} {538}},\ \bibinfo {pages} {110898} (\bibinfo
  {year} {2020})}\BibitemShut {NoStop}%
\bibitem [{\citenamefont {Ren}\ \emph {et~al.}(2021)\citenamefont {Ren},
  \citenamefont {Long},\ and\ \citenamefont {Deng}}]{ren2021new}%
  \BibitemOpen
  \bibfield  {author} {\bibinfo {author} {\bibfnamefont {B.}~\bibnamefont
  {Ren}}, \bibinfo {author} {\bibfnamefont {Z.}~\bibnamefont {Long}},\ and\
  \bibinfo {author} {\bibfnamefont {R.}~\bibnamefont {Deng}},\ }\bibfield
  {title} {\bibinfo {title} {A new criterion for predicting the glass-forming
  ability of alloys based on machine learning},\ }\href@noop {} {\bibfield
  {journal} {\bibinfo  {journal} {Computational Materials Science}\ }\textbf
  {\bibinfo {volume} {189}},\ \bibinfo {pages} {110259} (\bibinfo {year}
  {2021})}\BibitemShut {NoStop}%
\end{thebibliography}%

\section*{Supplementary Information}
Supplementary Notes

Supplementary Figure S1

Supplementary Tables S1-S3

\clearpage
\pagebreak
\beginsupplement
\pagestyle{plain}
\onecolumngrid

\begin{center}
  \huge {Supplementary Information for}
  \bigskip

  \large \textbf{Inverse Materials Design by Large Language Model-Assisted Generative Framework} \\ 

  \bigskip
  \bigskip

Yun Hao, Che Fan, Beilin Ye, Wenhao Lu, Zhen Lu, Peilin Zhao, Zhifeng Gao, Qingyao Wu, Yanhui Liu, and Tongqi Wen
\end{center}

\clearpage

\makeatletter

\titleformat{\subsubsection}
{\large}{\thesubsubsection}{1em}{{#1}}

\titleformat{\subsection}
{\bfseries\large}{\thesubsection}{1em}{{#1}}

\titleformat{\section}
{\bfseries\Large}{\thesection}{1em}{{#1}}

\section{Supplementary Notes}

\subsection{Relevant formulas in inverse design}

To quantify the similarity between two probability distributions, $P$ and $Q$, we employ three key metrics: Kullback–Leibler (KL) divergence $D_{\mathrm{KL}}(P \| Q)$, Jensen–Shannon (JS) divergence $D_{\mathrm{JS}}(P \| Q)$, and the Wasserstein distance.\\
KL divergence~\cite{kullback1951information} measures how one probability distribution diverges from a reference distribution. It is defined as:
\begin{equation}\label{eq:kl}
 D_{\mathrm{KL}}(P \| Q)=\int_{-\infty}^\infty P(x) \log \left(\frac{P(x)}{Q(x)}\right) \mathrm{d} x
\end{equation}
\\
JS divergence~\cite{tishby2000information} provides a symmetric and smoothed measure of the difference between two distributions by averaging their KL divergences with respect to their mixture distribution M:
\begin{equation}\label{eq:js}
    D_{\mathrm{JS}}(P \| Q)=\frac{1}{2} D(P \| M)+\frac{1}{2} D(Q \| M),
\end{equation}
where $M = \dfrac{1}{2}\left( P + Q \right) $ is a mixture distribution of $P$ and $Q$ . \\
\\
Wasserstein distance~\cite{vaserstein1969markov} measures the optimal transport cost required to morph one distribution into another::

\begin{equation}\label{eq:wd}
d_{w}(P,Q) =  \inf_{\gamma \in \Pi(P, Q)} \mathbb{E}_{(x, y) \sim \gamma} \left [ \left | x - y\right | \right]
\end{equation}

Here, \( \Pi(P, Q) \) represents the set of all joint distributions \( \gamma(x, y) \) whose marginals are \( P \) and \( Q \), respectively. Intuitively, \( \gamma \) describes the most efficient way to transport ``mass" from $P$ to $Q$, while \( ||x - y|| \) represents the distance between corresponding points, typically measured using the Euclidean metric.

\subsection{Relevant formulas in classification tasks\label{formulas:classification}}
To evaluate the performance of our classification model, we categorize predictions into four groups within the test set: true positives ($TP$) for correctly predicted positive cases, true negatives ($TN$) for correctly predicted negative cases, false positives ($FP$) for incorrectly predicted positive cases, and false negatives ($FN$) for incorrectly predicted negative cases. 
We assess classification accuracy using three key metrics: precision, recall, and $F^1_\text{score}$:

Precision measures the proportion of correctly identified positive cases out of all predicted positives:
\begin{equation}
    \text{precision} = \frac{TP}{TP+FP}
\end{equation}\label{eq:classfication1}

Recall quantifies the proportion of actual positive cases that were correctly identified:
\begin{equation}
    \text{recall} = \frac{TP}{TP+FN}
\end{equation}\label{eq:classfication2}

The $F^1_\text{score}$ score provides a balanced measure of precision and recall, calculated as their harmonic mean:
\begin{equation}
    F^1_\text{score} = 2*\frac{\text{precision}*\text{recall} }{\text{precision}+\text{recall} }
\end{equation}\label{eq:classfication3}
This metric ensures a robust evaluation of the model, particularly when there is an imbalance between positive and negative samples.

\subsection{Relevant formulas in regression tasks}

To evaluate the performance of the regression model, we use three key metrics: Root Mean Square Error (RMSE), Mean Absolute Error (MAE), and the Coefficient of Determination ($R^2$). These metrics quantify the accuracy of predicted values $\hat{y}_i$  compared to actual observations  $y_i$  over a dataset of $n$ samples.

Root Mean Square Error (RMSE): RMSE measures the average magnitude of prediction errors, giving higher weight to larger errors due to squaring:
\begin{equation}
\mathrm{RMSE}=\sqrt{\frac{1}{n}\sum_{i=1}^n\left(y_i-\hat{y}_i\right)^2}
\end{equation}

Mean Absolute Error (MAE): MAE represents the average absolute difference between predicted and actual values, providing a more interpretable measure of error:
\begin{equation}   
\text{MAE} = \dfrac{1}{n} \sum_{i=1}^{n} |y_i - \hat{y_i}|
\end{equation}

Coefficient of Determination ($R^2$): The $R^2$ score (Nash–Sutcliffe model efficiency coefficient~\cite{nash1970}) quantifies how well the regression model explains the variance in the data. It is defined as:
\begin{equation} 
R^2=1-\frac{S S_{\mathrm{res}}}{S S_{\mathrm{tot}}}
\end{equation}
where Residual Sum of Squares ($SS_{\mathrm{res}}$) measures the total discrepancy between the observed and predicted values:
\begin{equation}
SS_{\mathrm{res}}=\sum_{i=1}^n\left(y_i-\hat{y}_i\right)^2
\end{equation}
Total Sum of Squares ($SS_{\mathrm{tot}}$) quantifies the overall variance in the dataset by comparing each observed value to the mean $\bar{y}$, where:
\begin{equation}
    \bar{y} = \frac{1}{n} \sum_{i=1}^{n} y_i
\end{equation}

\begin{equation}
SS_{\mathrm{tot}}=\sum_{i=1}^n\left(y_i-\bar{y}\right)^2
\end{equation}

An $R^2$ score close to 1 indicates a strong correlation between predictions and actual values, while a lower $R^2$ suggests a weaker fit.

\newpage

\clearpage
\section{Supplementary Figures}

\begin{figure*}[h!]
	\begin{center}
		\centering
		\includegraphics[width=7in]{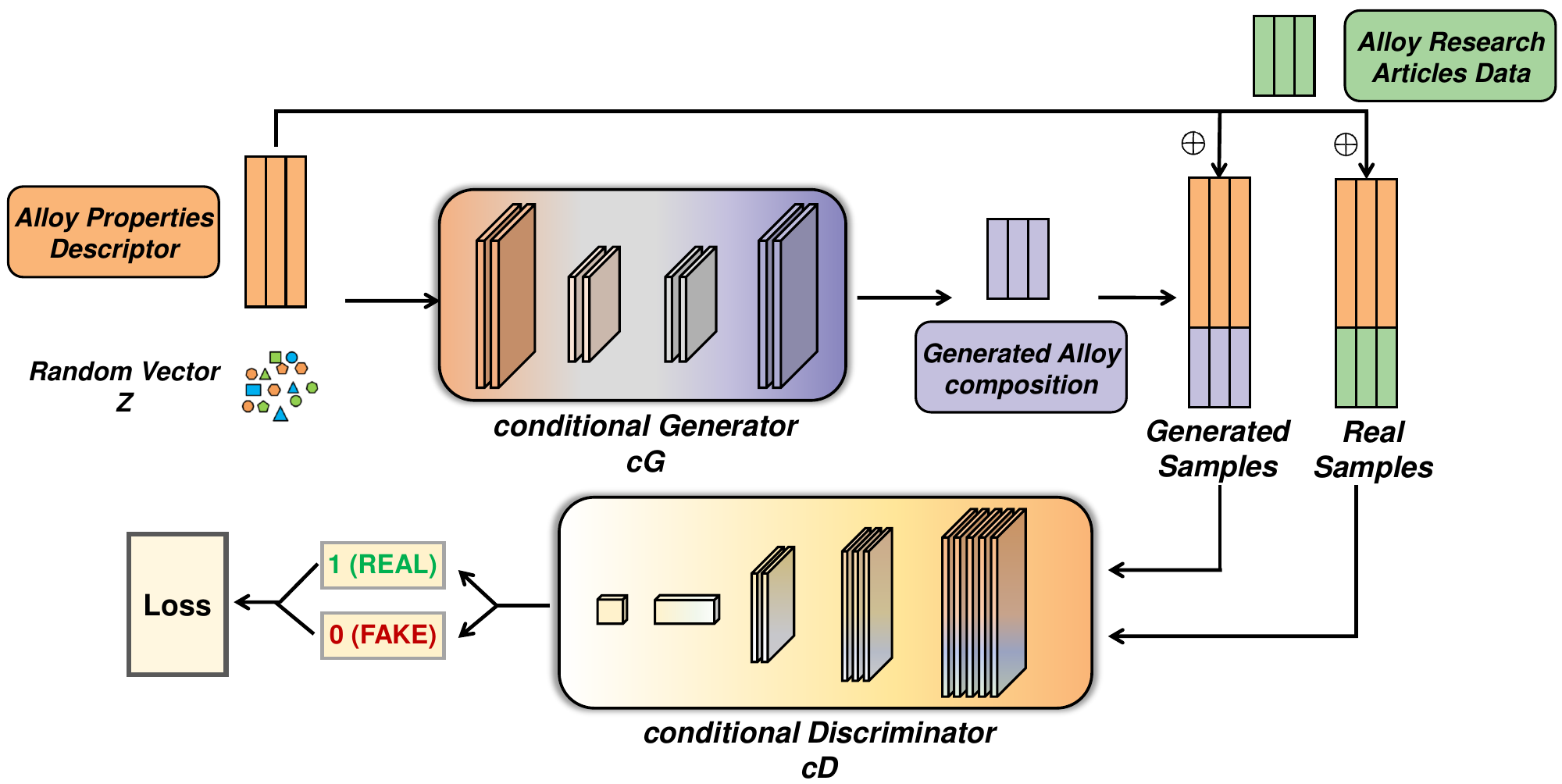}
	\end{center}
	\caption{Schematic of the CGAN framework for inverse alloy design. The conditional generator (cG) takes an Alloy Properties Descriptor and a random noise vector (Z) as inputs to generate synthetic alloy compositions. The conditional discriminator (cD) evaluates the generated compositions by distinguishing between two inputs: (i) synthetic alloy compositions concatenated with the Alloy Properties Descriptor, and (ii) real alloy compositions extracted from literature, also concatenated with the Alloy Properties Descriptor. The discriminator provides feedback to the generator, guiding it to produce increasingly realistic alloy compositions that satisfy the specified property constraints.}
	\label{fig:cgan_chart}
\end{figure*}

\clearpage
\section{Supplementary Tables}

\begin{table}[htbp]
\setlength{\tabcolsep}{8mm}{
\caption{Thermophysical parameters for glass-forming ability. Summary of the key thermophysical parameters used to assess glass-forming ability in this study.}
\label{tab:descriptor_si}
\begin{tabular}{lllll}
\toprule
No. & Parameter & Formula & Proposed by & Year \\
\midrule
1 & $\Delta T_{x}$ & $\Delta T_{x} = T_{x} - T_{g}$ & Inoue A \cite{inoue1991mg}& 1991\\
2 & $T_{rg}$ & $T_{rg} = T_{g}/T_{l}$ & Lu ZP \cite{lu2000correlation}& 2000\\
3 & $\gamma$ & $\gamma = T_{x}/(T_{g}+T_{l})$ & Lu ZP \cite{lu2002new}& 2002\\
4 & $\Delta T_{rg}$ & $\Delta T_{rg} = \Delta T_{x}/(T_{l}-T_{g})$ & Xiao XS \cite{xiao2004influence}& 2004\\
5 & $\alpha$ & $\alpha = T_{x}/T_{l}$ & Mongal K \cite{mondal2005parameters}& 2005\\
6 & $\beta_{1}$ & $\beta_{1} = T_{x}/T_{g}+T_{g}/T_{l}$ & Mongal K \cite{mondal2005parameters}& 2005\\
7 & $\delta$ & $\delta = T_{x}/(T_{l}-T_{g})$ & Chen QJ \cite{chen2006new}& 2006\\
8 & $\gamma_{m}$ & $\gamma_{m} = (2T_{x}-T_{g})/T_{l}$ & Du XH \cite{du2007new} & 2007\\
9 & $\varphi$ & $ \varphi=T_{g}/T_{l}(\Delta T_{x}/T_{g})^{0.143} $ & Fan GJ \cite{fan2007new}& 2007\\
10 & $\xi $ & $\xi = T_{g}/T_{l}+\Delta T_{x}/T_{x}$ &Du XH \cite{du2008new}& 2008\\
11 & $\beta_{2} $ & $\beta_{2} = T_{x}T_{g}/(T_{l}-T_{x})^2$ & Yuan ZZ \cite{yuan2008new}& 2008\\
12 & $\omega_{1} $ & $\omega_{1} =T_{g}/T_{x}-2T_{g}/(T_{g}+T_{l}) $ & Long ZL \cite{long2009new}& 2009\\
13 & $\omega_{3} $ & $\omega_{3} =(T_{l}+T_{x})/(T_{l}-T_{x})*T_{l}/T_{x} $ & Ji XL \cite{ji2009thermodynamic}& 2009\\
14 & $\theta $ & $\theta = [(T_{g}+T_{x})/T_{l}*(T_{x}-T_{g})/T_{l}]^{0.0728}$ & Zhang GH \cite{zhang2009criterion}& 2009\\
15 & $\omega_{2} $ & $\omega_{2} = T_{g}/(2T_{x}-T_{g})-T_{g}/T_{l}$ & An LX \cite{2009Correlations}& 2009\\
16 & $\gamma_c $ & $\gamma_c = (3T_{x}-2T_{g})/T_{l}$ & Guo S \cite{guo2010new}& 2010\\
17 & $\beta^{'} $ & $\beta^{'} = T_{g}/T_{x}-T_{g}/(1.3T_{l})$ & Dong BS \cite{dong2011new}& 2011\\
18 & $\omega_{4} $ & $\omega_{2} = (2T_{x}-T_{g})/(T_{l}+T_{x})$ & Blyskum P \cite{blyskun2015relation}& 2015\\
19 & $G_p $ & $G_p = [(T_{g}(\Delta T_{x})/(T_{l}-T_{x})^2]$ & Tripathi MK \cite{tripathi2016evolution}& 2016\\
20 & $\chi^{'} $ & $\chi^{'} = \Delta T_{x}/(T_{l}-T_{x})[T_{x}/(T_{l}-T_{x})]^{1.47}$ & Long ZL \cite{long2018new}& 2018\\
21 & $\gamma_{n}$ & $\gamma_{n} = (5T_{x}-3T_{g})/T_{l} $ & Xiong J \cite{xiong2019machine}& 2019\\
22 & $\nu$ & $\nu = T_{x}*T_{g}*\Delta T_{x}/(T_{l}-T_{x})^3 $ & Deng RJ \cite{deng2020critical}& 2020\\
23 & $k$ & $k = T_{x}*T_{g}*T_{l}*\Delta T_{x}/(T_{l}-T_{x})^4$ & Ren BY \cite{ren2021new}& 2021\\
\bottomrule
\end{tabular}}
\end{table}

\begin{table}[htbp]
\setlength{\tabcolsep}{8mm}{
\caption{Performance comparison of AlloyGAN with previous inverse design methods in regression tasks. AlloyGAN demonstrates superior predictive accuracy for $T_{rg}$ and $\gamma$ as regression targets, achieving the lowest RMSE and MAE values while significantly improving $R^2$ scores.}
\label{tab:comparasion}
\begin{tabular}{cllll}
\toprule
Selected Features & Methods& RMSE &  MAE & $R^2$\\
\midrule
\multirow{2}{*}{$T_{rg}$}& Zhou, et al \cite{zhou2023generative}&0.0265&0.0160&0.1618\\
 & AlloyGAN (Ours) &\textbf{0.0155 $\downarrow$}&\textbf{0.0109$\downarrow$}&\textbf{0.7429$\uparrow$}\\
\midrule
\multirow{2}{*}{$\gamma$}& Zhou, et al \cite{zhou2023generative}&0.0126&0.0096&0.5393\\
&AlloyGAN (Ours) &\textbf{0.0074$\downarrow$}&\textbf{0.0058$\downarrow$}&\textbf{0.8030$\uparrow$}\\
\bottomrule
\end{tabular}}
\end{table}

\begin{table}[h]
    \centering
    \renewcommand{\arraystretch}{1.5}
    \setlength{\tabcolsep}{8mm} 
    \caption{Candidate alloy compositions and their predicted thermodynamic parameters. AlloyGAN generates listed compositions for good glass-forming ability with $T_{rg}>0.6$ and $\gamma>0.4$.}
    \label{tab:alloy_data}
    \begin{tabular}{lccccc}
        \hline
        \textbf{Cu-based Alloys} & $T_g$ (K) & $T_x$ (K) & $T_l$ (K) & $T_{rg}$ & $\gamma$ \\
        \hline
        Cu$_{51.5}$Zr$_{45.2}$Ti$_{2}$Hf$_{1.3}$& 630 & 650 & 957 & 0.66 & 0.41 \\
        Cu$_{51.2}$Ti$_{21.1}$Ni$_{14.6}$Hf$_{9.7}$Zr$_{3.4}$& 611 & 655 & 961 & 0.64 & 0.42 \\
        Cu$_{56.4}$Ti$_{15.7}$Zr$_{9.9}$Ni$_{6.9}$Hf$_{11.1}$& 720 & 752 & 1157 & 0.62 & 0.40 \\
        Cu$_{51.8}$Zr$_{48.2}$ & 600 & 645 & 995 & 0.60 & 0.40 \\
        Cu$_{51.6}$Zr$_{47.9}$Hf$_{0.5}$ & 689 & 743 & 1142 & 0.60 & 0.41 \\
        \hline
        \textbf{Fe-based Alloys} & $T_g$ (K) & $T_x$ (K) & $T_l$ (K) & $T_{rg}$ & $\gamma$ \\
        \hline
        Fe$_{78.1}$B$_{21.5}$Si$_{0.2}$Co$_{0.2}$ & 905 & 933 & 1424 & 0.64 & 0.40 \\
        Fe$_{75.4}$B$_{24.5}$Si$_{0.1}$ & 896 & 941 & 1414 & 0.63 & 0.41 \\
        Fe$_{76}$B$_{23.8}$Co$_{0.2}$ & 920 & 941 & 1483 & 0.62 & 0.39 \\
        Fe$_{80.6}$B$_{19.2}$Si$_{0.1}$Co$_{0.1}$ & 915 & 944 & 1488 & 0.61 & 0.39 \\
        Fe$_{75}$B$_{24.9}$Co$_{0.1}$ & 752 & 799 & 1226 & 0.61 & 0.40 \\
        \hline
        \textbf{Ti-based Alloys} & $T_g$ (K) & $T_x$ (K) & $T_l$ (K) & $T_{rg}$ & $\gamma$ \\
        \hline
        Ti$_{47.9}$Cu$_{37.9}$Zr$_{8.1}$Sn$_{3.4}$Ni$_{2.7}$ & 623 & 644 & 953 & 0.65 & 0.41 \\
        Ti$_{45.7}$Cu$_{41.1}$Zr$_{7.6}$Sn$_{3.2}$Ni$_{2.4}$ & 720 & 752 & 1157 & 0.62 & 0.40 \\
        Ti$_{48.9}$Cu$_{35.5}$Zr$_{9.9}$Sn$_{3}$Ni$_{2.7}$ & 600 & 645 & 995 & 0.60 & 0.40 \\
        \hline
        \textbf{Zr-based Alloys} & $T_g$ (K) & $T_x$ (K) & $T_l$ (K) & $T_{rg}$ & $\gamma$ \\
        \hline
        Zr$_{60.8}$Cu$_{30.5}$Al$_{4.3}$Ag$_{2.7}$Ni$_{1.7}$& 650 & 700 & 951 & 0.68 & 0.44 \\
        Zr$_{58.6}$Cu$_{30.7}$Al$_{5.3}$Ag$_{3.9}$Ni$_{1.5}$& 670 & 712 & 1029 & 0.65 & 0.42 \\
        Zr$_{58.1}$Cu$_{32.8}$Al$_{5.9}$Ag$_{3.2}$& 653 & 733 & 1029 & 0.63 & 0.44 \\
        Zr$_{62.1}$Cu$_{31.4}$Al$_{5.1}$Ni$_{1.4}$& 699 & 769 & 1121 & 0.62 & 0.42 \\
        Zr$_{62}$Cu$_{29.6}$Al$_{4.2}$Ag$_{2.3}$Ni$_{1.9}$& 698 & 765 & 1125 & 0.62 & 0.42 \\
        \hline
    \end{tabular}
\end{table}

\end{document}